\documentclass[rmp]{revtex4}
\usepackage{amssymb,amsfonts,amsmath}
\usepackage{setspace}
\usepackage{hyperref}
\usepackage{graphicx}
\usepackage{amsmath}
\usepackage{natbib}

\newcommand{\beast}[1]{\emph{#1}}
\newcommand{\gene}[1]{\emph{#1}}
\newcommand{\xs}{\chi}

\newcommand{\rs}{\tilde{r}}
\newcommand{\dss}{{\tilde{s}}}
\newcommand{\mut}{U}
\newcommand{\x}{x}

\newcommand{\mfit}{\bar{\x}}
\newcommand{\rec}{r}
\newcommand{\ds}{s}

\newcommand{\be}{\begin{equation}}
\newcommand{\ee}{\end{equation}}
\newcommand{\bea}{\begin{eqnarray}}
\newcommand{\eea}{\end{eqnarray}}

\newcommand{\gr}{\theta}
\newcommand{\Thc}{\Theta_c}
\newcommand{\Thi}{\Theta_0}

\newcommand{\ti}{\tau_0}

\newcommand{\Ne}{N_e}
\newcommand{\Tfix}{T_{fix}}
\newcommand{\pfix}{p_{fix}}
\newcommand{\Tfp}{T_{fp}}
\newcommand{\Tcoal}{T_C}
\newcommand{\Pnx}{P_{S}}
\newcommand{\Pallele}{f}
\newcommand{\Pnumber}{p}

\bibpunct{(}{)}{;}{author-year}{}{,} 

\begin{document}
\title{Genetic Draft and Quasi-Neutrality in Large Facultatively Sexual Populations.}
\author{R.~A.~Neher${}^{*}\dagger$ and B.~I.~Shraiman${}^{*\ddagger}$\\
${}^{*}$Kavli Institute for Theoretical Physics, \\${}^{\ddagger}$Department of Physics, University of
California, Santa Barbara, CA 91306 and \\${}^{\dagger}$Max-Planck Institute for Developmental Biology, T\"ubingen, 72070, Germany}
\date{\today}
\maketitle

\bibliographystyle{genetics}
\section*{Abstract}
Large populations may contain numerous simultaneously segregating polymorphisms subject to natural selection. Since selection acts on individuals whose fitness depends on many loci, different loci affect each other's dynamics. This leads to stochastic fluctuations of  allele frequencies above and beyond genetic drift - an effect known as \emph{genetic draft}. Since recombination disrupts associations between alleles, draft is strong when recombination is rare.
Here, we study a facultatively outcrossing population in a regime where the frequency of out-crossing and recombination, $\rec$, is small compared to the characteristic scale of fitness differences $\sigma$. In this regime, fit genotypes expand clonally, leading to large fluctuations in the number of recombinant offspring genotypes. The power law tail in the distribution of the latter makes it impossible to capture the dynamics of draft by an effective neutral model. Instead, we find that the fixation time of a neutral allele increases only slowly with the population size but depends sensitively on the ratio $\rec/\sigma$. The efficacy of selection is reduced dramatically and alleles behave ``quasi-neutrally" even for $N\ds\gg 1$, provided that $|s|< s_c$, where $s_c$ depends strongly on $\rec/\sigma$, but only weakly on population size $N$. In addition, the anomalous fluctuations due to draft change the spectrum of (quasi)-neutral alleles from $\Pallele(\nu)\sim \nu^{-1}$, corresponding to drift, to $\sim \nu^{-2}$. Finally, draft accelerates the rate of two step adaptations through deleterious intermediates.

\section*{Introduction}
The genetic diversity of a population is determined by mutation, selection, recombination and genetic drift, i.e.~the stochasticity inherent in reproduction. Understanding how genetic diversity depends on these elements of evolutionary dynamics is central to population genetics, since it allows to make inferences about the past history and to predict how rapidly populations can adapt. 


Population genetic inference focuses on the diversity at putatively neutral sites and assumes that the history of these sites is described by the neutral ``coalescent'' \cite{Kingman:1982p28911}. Coalescent theory models the genealogy of asexual organisms or non-recombining segments of a genome by positing that lineages merge at random, backwards in time, due to common ancestry. Under this assumption, the mean time to the most recent common ancestor, $\Tcoal$, of the extant $N$ individuals, is $2N$ generations. The coalescence time scale is very important, since the genetic diversity of the population is given by the number of mutations that occur in all lineages descending from the most recent common ancestor of the population. Genetic diversity is therefore controlled by $\Tcoal$ and hence, under the assumption of neutral evolution, proportional to $N$. (Coalescent theory has been extended to weak selection \citep{Krone:1997p19666} and recombination \citep{Hudson:1983p37019,Griffiths:1996p25579}.) 

However, the prediction that neutral genetic diversity is proportional to $N$ is at odds with observations: population sizes of different organisms differ by many orders of magnitude, while genetic variation among organisms is fairly constant  \citep{Lewontin1974}. To resolve this `paradox of variation', \citet{Smith:1974p34217} suggested that selection acting on linked loci might reduce diversity at a neutral locus. Rapid fixation of a novel mutation at a linked locus will perturb the allele frequencies. These perturbations can bring alleles to fixation and, more generally, reduce the coalescence time and hence the average genetic diversity \citep{Kaplan:1989p34931,Barton:1998p28270,Gillespie:2001p9636}. Such ``hitch-hiking'' of neutral alleles on linked selected loci will dominate over genetic drift in large populations. Since hitch-hiking leads to larger perturbations for more closely linked loci, one expects genetic variation to correlate with the recombination rate, as is indeed observed in \beast{Drosophila} \citep{Begun:1992p34015,Stephan:1992p35194,Andolfatto:2001p33951,Sella:2009p26729}.

A related effect was described earlier by \citet{Hill:1966p21029}, who studied the reduction in the fixation probability of a novel beneficial mutation because of selection acting at a linked locus. This effect is commonly known as Hill-Robertson interference \citep{Felsenstein:1974p23937}. Hitchhiking and Hill-Robertson interference are different aspects of the same phenomenon, one focusing on the effects of linked selection on genetic diversity, the other on the efficacy of selection. 
While hitchhiking and Hill-Robertson effects are primarily associated with positive selection for novel alleles, purifying selection against deleterious mutations also affects genetic diversity. The elimination of (neutral) alleles linked with deleterious mutations is known as background selection. The lower the recombination rates are, the larger is the target for linked deleterious mutations, resulting in stronger background selection \citep{Charlesworth:1993p36005,Hudson:1995p18197,Nordborg:1996p18149}. 


Most models used to study Hill-Robertson and hitch-hiking effects between positively selected mutations consider only two loci. Deleterious mutations, however, are expected to be much more frequent, and background selection models typically consider many mutations with small deleterious effects. A systematic study of the effect of interference between many weakly selected sites in a mutation/selection/drift equilibrium was presented by \citet{McVean:2000p19278}, who used computer simulations of a model of codon bias evolution. They showed that linkage dependent interference between a large number of weakly selected sites has substantial effects on genetic diversity, fixation probability of mutations and the degree of adaptation measured as the frequency of preferred codons. This and subsequent computational studies reinforced the understanding that the Hill-Roberson effect reduces the effectiveness of selection and made clear that a quantitative understanding of Hill-Robertson effects in multi-locus systems requires an analysis that goes beyond two locus models \citep{Comeron:2002p18509,Seger:2010p39561}, see \citep{Comeron:2008p18004} for a recent review.

It is common to interpret the effect of linked selection in terms of increased variance in offspring number. In this interpretation, linked selection can be thought of as a stochastic force analogous to genetic drift and is often referred to as \emph{genetic draft} -- a term coined by \citet{Gillespie:2000p28513}.
Following \citet{Hill:1966p21029} and \citet{Felsenstein:1974p23937}, the increased variance in offspring number is often captured by a reduction in the ``effective population" size,  $\Ne$, in a neutral model (which means enhanced drift and accelerated coalescence). It has, however, been noted that a rescaled neutral model does not consistently explain all observables \citep{Charlesworth:1993p36005,Braverman:1995p34932,Fay:2000p35077,McVean:2000p19278,Barton:2004p34826,Seger:2010p39561}, and that different effective population sizes need to be defined depending on the question  and time scale of interest \citep{Ewens_2004,Karasov:2010p35377}. Furthermore, the dependence of $\Ne$ on the actual population size and other relevant parameters is not understood \citep{Wiehe:1993p37333,Gillespie:2000p28513,Lynch:2007p38185}.

Here, we provide analytic results on the effect of draft in a stochastic multi-locus evolution model. Instead of a mutation/selection equilibrium considered in \citet{McVean:2000p19278}, our focus here is a continuously adapting and facultatively sexual population, like HIV in coevolution with the host's immune system. Our model and its relation to the biology of HIV are described in more detail below. The scope of the model, however, extends beyond HIV and is equally applicable to scenarios where background selection is dominant. Many important and well studied organisms such as influenza, yeast and plants are facultatively sexual. Rice, for example, is a partly selfing species and strong selection has acted during its domestication \citep{Caicedo:2007p36637}. While dominance effects can render the selfing of diploid organisms more complicated than facultatively sexual propagation of haploid organisms \citep{Charlesworth:1991p39183,Kelly:2000p39103}, our analysis should still provides a null model on top of which dominance effects can be investigated.

Using computer simulations of an adapting population, we first demonstrate how quantities such as the coalescence time, the fixation probability of beneficial or deleterious mutations, as well as the allele frequency spectra, depend on the rate of outcrossing relative to selection. We also show that our {\it in silico} observations cannot be described by a neutral model with a reduced effective population size. This is because single genotypes can, through clonal expansion, give rise to a wildly fluctuating number of recombinant genotypes. The distribution is so broad, that its variance diverges, which in turn makes an effectively neutral diffusion limit impossible. To provide an analytic understanding of the simulation results, we use a branching process model that allows us to study the stochastic dynamics of novel mutations (neutral, beneficial and deleterious) as they spread through the population. We calculate fixation probabilities and the typical time to fixation, $\Tfix$, (and more generally, the probability of attaining $n$ copies after time $T$) for a new mutant allele, making explicit the dependence on the rate of recombination, fitness variance and the population size. 
An important consequence of genetic draft is a qualitatively different frequency spectrum of rare alleles, which we also calculate analytically. 
Finally, we will show that empirical HIV allele frequency spectra are in agreement with our theoretical prediction, confirming the relevance of our model to the dynamics of HIV adaptation.

\section*{Model}
\label{sec:model}

Our model is inspired by the intra-patient evolution of HIV.
We will first outline briefly the biology of HIV, and then describe our computational model and the branching process approximation used to study the phenomena analytically.

\subsection*{Intra-Patient evolution of HIV}
After successful infection, the virus proliferates quickly to levels of more than $10^6$ viral genomes per milliliter blood. This acute phase lasts for several weeks, after which the immune system of the host reduces the number of viral genomes per milliliter blood to roughly $100-10^4$. The following chronic infection can last for many years and is largely asymptomatic up to the onset of AIDS. However, even in the chronic phase of infection the total number of viruses produced per day is estimated to be $\sim 10^{10}$ per ml, while the generation time is roughly two days \citep{Perelson:1996p23158,Markowitz:2003p39214}. The viral population is subject to clearance by cytotoxic T-lymphocyte (CTL) mediated cell death and anti-body mediated destruction of virus \citep{Paul2008}, which puts the virus under constant selection to change its proteins. CTL epitopes are found throughout the HIV genome \citep{LANLImmunology} and escape mutations are often selected for with coefficients of a few percent per generation \citep{Asquith:2006p28003}. Antibodies bind primarily to the products of the envelope gene. In response to antibody recognition, numerous escape mutants arise in the envelope gene and spread through the population driven by selection coefficients of one to a few percent \citep{Williamson:2003p26136,Neher:2010p32691}. Additional, even stronger, selection pressure is put on the viral population by drug treatment, resulting in rapid emergence of resistance during suboptimal therapy \citep{Larder:1989p28713,Hedskog}. Several hundred mutations are implicated in drug resistance \citep{Rhee:2003p24151}.

HIV carries two copies of its RNA genome ($10^4$ bps), from which one cDNA is produced and integrated into the host cell genome. Recombination in HIV occurs through frequent template switching of the reverse transcriptase in the process of cDNA \citep{Levy:2004p23309} production. In other words, the two genomes of the ``diploid" virion are combined to produce a ``haploid" cDNA, from which all the viral proteins and the next generations genomes are produced. The rate of recombination is limited by co-infection of host cells with several viruses, which is necessary to produce a heterozygous virus (see Figure~\ref{fig:hivrecomb}A). Estimates suggest an effective recombination rate of $\approx 1.5\times 10^{-5}$ per base per generation \citep{Neher:2010p32691}, corresponding to a co-infection rate $\leq 10\%$. Viruses therefore undergo clonal amplification most of the time, while different parts of the genome are only weakly linked in the event of ``outcrossing'' (heterozygote formation, followed by template switching).  

\subsection*{Computational model}
Based on the discussion above our model must include the following elements: a large population, selection at many polymorphic loci, a constant supply of beneficial mutations, and facultative mating with substantial reassortment in case of outcrossing. Models combining all of these elements have already been established in \citep{Rouzine:2005p17398,Neher:2010p30641,Rouzine:2010p33121} and a similar model will be used here.  In our simulation a (nearly) constant variance of fitness in the population, $\sigma^2$, is maintained by a constant supply of beneficial mutations. This results in a continuously adapting population, with a rate of adaptation given approximately by the variance in fitness. 
Of course, one does not expect constant selection to persist indefinitely (see \citep{Mustonen:2010p36757} for a discussion of evolution in changing environments): steady conditions on the time scale of fixation of a single allele would suffice for the validity of our analysis.

To simulate population dynamics with the ingredients described above, we implemented a discrete time (Fisher-Wright) dynamics of individuals with haploid genomes and a large number of loci which additively define the (log-)fitness $\x$ of an individual. Each individual contributes a Poisson distributed number of gametes to the next generation with mean $\exp(\x-\mfit - \alpha)$, where $\mfit$ is the current mean fitness and $\alpha$ is adjusted to keep the population size approximately constant at $N$. To implement facultative mating, a fraction $\rec$ of the gametes are paired at random, the alleles at all of their loci reassorted, and the two resulting recombinant offspring are placed into the next generation. The remaining $1-\rec$ fraction of gametes is placed unchanged into the next generation. New beneficial mutations with effect size $\ds_0$ ($N\ds_0\gg 1$) are introduced into a randomly chosen individual at rate $N\mut_b\gg 1$. After equilibration, the fitness distribution in the population is approximately Gaussian with a nearly constant variance and steadily increasing mean fitness.  Into this adapting population wave, we introduce additional mutations, neutral, deleterious or beneficial, and track their dynamics. In particular, we record after what time these mutations reach certain frequency thresholds and the fitness of genotypes on which successful mutations originate. We also measure the allele frequency spectra and the cumulative number of individuals that carried a particular mutation before it goes extinct. A further, and more detailed, description of the implementation is given in Appendix \ref{sec:app_model_implementation}.

\begin{figure}[htp]
\begin{center}
  \includegraphics[width=0.7\columnwidth]{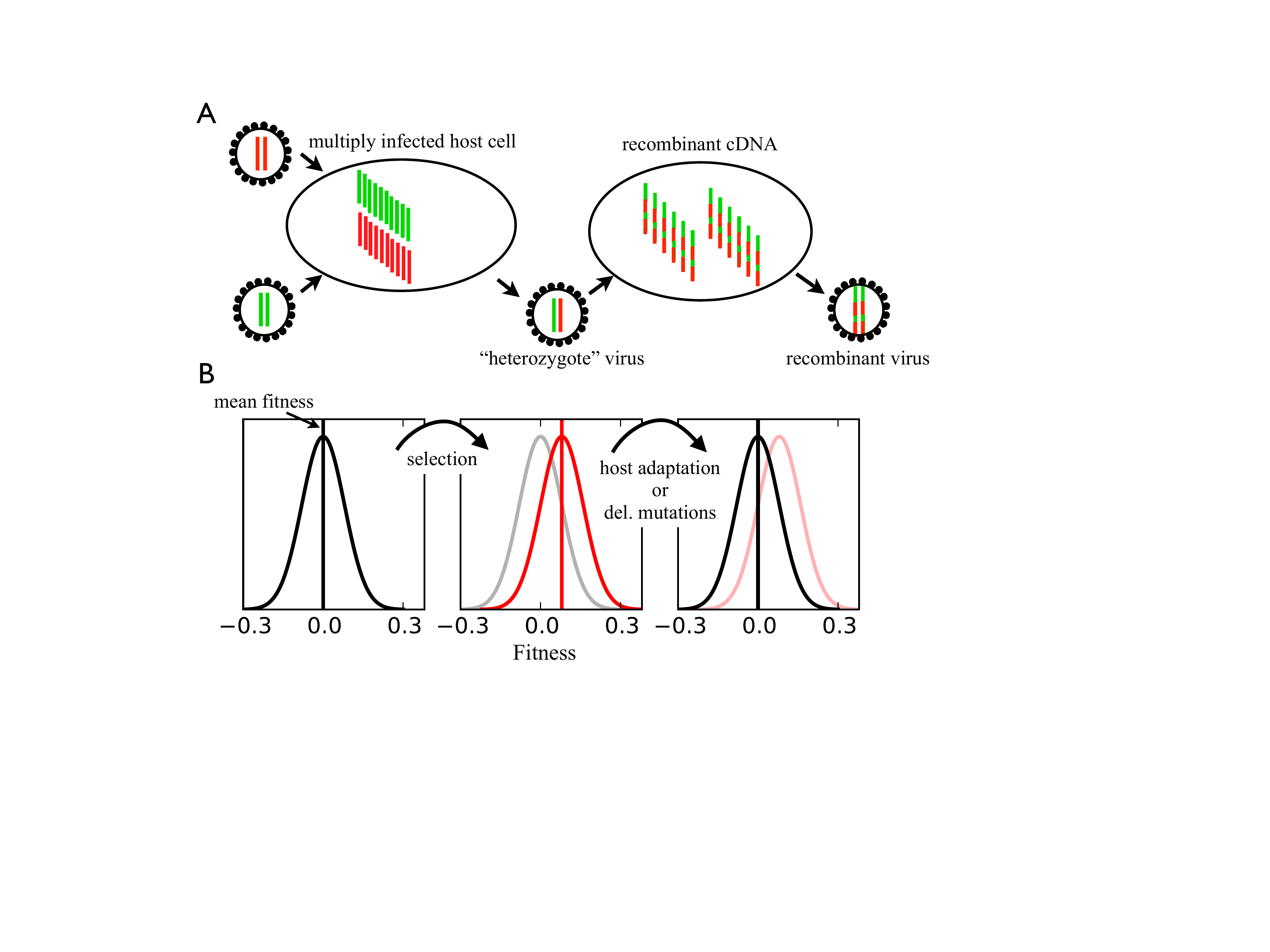}
  \caption[labelInTOC]{Panel A illustrates the process of recombination in HIV. Diversifying recombination occurs only in ``heterozygote'' virions, which require multiple infections of the host cell in the previous generation. This is rare, leading to an effectively facultative sexual population. Panel B illustrates the dynamics of the fitness distribution. Selection and mutation are shown as separate steps, for simplicity. Selection, acting between the left and center panel, results in a shift of the fitness distribution towards higher fitness. In a deleterious mutation/selection balance, this increase in fitness is compensated by deleterious mutations occurring during the step leading from the center to the right panel. If deleterious mutations are not the dominant force, but continuous adaptation to a coevolving host is modeled, the adaptation of the host counter balances the fitness advance of the population. }
  \label{fig:continuous_evo}
  \label{fig:hivrecomb}
\end{center}
\end{figure}

Our focus here is on rapidly adapting populations with many concurrent sweeps responsible for the fitness variance $\sigma^2$. HIV, as any other organism, also suffers from deleterious mutations which vastly outnumber the beneficial mutations. Deleterious mutations also contribute to fitness variation and increase $\sigma^2$. For the present investigation, it does not matter whether fitness variation is due to sweeping beneficial mutations, deleterious mutations subject to purifying selection, or a combination of the two. The two extreme scenarios are illustrated in Fig.~\ref{fig:continuous_evo}B. We shall see that the fate of novel mutations depends on how the fitness of clones changes with respect to the mean fitness of the population. It is irrelevant whether this fitness difference decreases because the mean fitness increases, or because the fitness of the clone decreases due to deleterious mutations.

\subsection*{Mathematical model}
\label{sec:BP_model}
To understand the results of the computer simulations, we analyze analytically an idealized model, which describes the propagation of a new allele (under selection, drift and occasional recombination onto a different genetic background) within a population with a fitness distribution approximated by a Gaussian traveling wave. This is justified if many beneficial mutations are sweeping through the population at the same time (which requires $N\mut_b\gg 1$ and is true for HIV) and is supported by our numerical simulations. Gaussian fitness distributions are expected in large asexual populations \citep{Tsimring:1996p19688,Rouzine:2003p33590,Desai:2007p954} and in the context of background selection (the deterministic equilibrium distribution in case of multiplicative selection is Poisson \citep{Haigh:1978p37141}, which is well approximated by a Gaussian in the limit of interest). Recombination stabilizes the Gaussian form if outcrossing rates are much larger than the inverse of the coalescence times \citep{Rouzine:2005p17398,Cohen:2005p5007}, see Fig.~S1 in the supplement. 
Selection moves this Gaussian distribution towards higher fitness at a rate $v$ equal to the variance in fitness $\sigma^2$ (Fisher's ``Fundamental Theorem" of natural selection). 

\emph{Branching process approximation.}
The fate of a novel allele is decided mainly at early times when the allele is rare and the finite population size constraint can be neglected. In this case the dynamics of the novel allele is governed by a stochastic branching process which in addition to the birth/death events also accounts for the recombination process which transfers the novel allele to different genetic backgrounds with different fitness $\x'$ \citep{Barton:1995p2811,Neher:2010p30641}.  Within the branching process approximation the probability of finding $n$ copies of the allele (with effect $\ds$ on fitness) anywhere in the population at time $T$, given that there were $k$ copies on background genotypes with  fitness $\x$ at time $t$, obeys the equation (see Supplement)
\begin{equation}
\begin{split}
\label{eq:BME}
-\partial_t \Pnumber(n,T|k,t,\x)=& -k(B+D+\rec)\Pnumber(n,T|k,t,\x)  +
kB \Pnumber(n,T|k+1,t,\x)+k D \Pnumber(n,T|k-1,t,\x)\\
&+rk\sum_{n'}\int d\x' K_{\x\x'}\Pnumber(n-n',T|k-1,t,\x)\Pnumber(n',T|1,t,\x') \ ,
\end{split}
\end{equation}
where $B=1+\x-\bar{\x}(t)+\ds$ is the birth rate, $D=1$ the rate at which individuals die (used to set the unit of time), and $\rec$ is the outcrossing rate. The first term describes the probability flux out of state $(k,t)$. The second term corresponds to a birth, which happens with rate $kB$ and in which case the final state $(n,T)$ is reached with probability $\Pnumber(n,T|k+1,t,\x)$. Analogously after a death, the probability of reaching $(n,T)$ is $\Pnumber(n,T|k-1,t, \x)$. The last term describes the process of outcrossing with the offspring fitness distribution parameterized by a recombination function $K_{\x\x'}$, which depends on the details of the model. 

In a model of outcrossing where offsprings are produced by mating two individuals at random and re-assorting all of their genes, the recombination kernel $K_{\x\x'}$, after averaging over the mating partner, is given by 
\begin{equation}
\label{eq:fullmodel}
K_{\x\x'} =
\sqrt{\frac{2}{3\pi\sigma^2}}e^{-\frac{2\left(\x'-\bar{\x}(t)-\frac{x-\bar{\x}(t)}{2}\right)^2}{3\sigma^2}}
\end{equation}
In this model, known as the \emph{infinitesimal model}, the fitness distribution of the offspring fitness is centered around half the fitness of the parent carrying the focal allele, measured relative to the instantaneous mean fitness $\bar{\x}(t)$ \citep{Bulmer_1980,Barton:2009p31030} (the model was referred to as the \emph{full recombination model} in \citep{Neher:2010p30641}). (Note that the variance of offspring fitness relative to parental mean is $\sigma^2/2$. The variance of $3\sigma^2/4$ in Eq. (\ref{eq:fullmodel}) is the result of averaging over the mate.) We assume that individuals are polygamous, i.e.~each offspring is the result of an independent mating event.

For most of the analytic calculations, we employ an even simpler recombination model, where offspring fitness is simply a random sample from the population, independent of the parents.
\begin{equation}
\label{eq:poolmodel}
K_{\x\x'} =
\frac{1}{\sqrt{2\pi\sigma^2}}e^{-\frac{(\x'-\bar{\x}(t))^2}{2\sigma^2}}
\end{equation}
In this \emph{communal recombination} model new offspring are assembled from
the genetic diversity in the entire population, i.e.~the novel genotype is drawn from a product distribution with the current allele frequencies in the population \citep{Barton:2009p31030,Neher:2010p30641}. Note, however, that this model does not lead to instantaneous elimination of correlations between loci (linkage disequilibrium), since only a fraction of the population outcrosses in a single generation. The infinitesimal and communal model have been shown to yield very similar results in \citet{Neher:2010p30641}. We will confirm this again by simulation of both models. Even though this model is a drastic idealization, it might resemble rather closely the recombination process in HIV. Since recombination in HIV depends on coinfection of T-cells with several viruses, most recombination occurs in centers with high concentrations of virus \citep{Jung:2002p23469}. In these centers, lineages might undergo several successive outcrossing events, effectively producing a cloud of offspring that contain genetic material from a large number of parents.

\section*{Results}
\label{sec:results}
\subsection*{Simulation results}
\begin{figure}[htp]
\begin{center}
  \includegraphics[width=0.45\columnwidth]{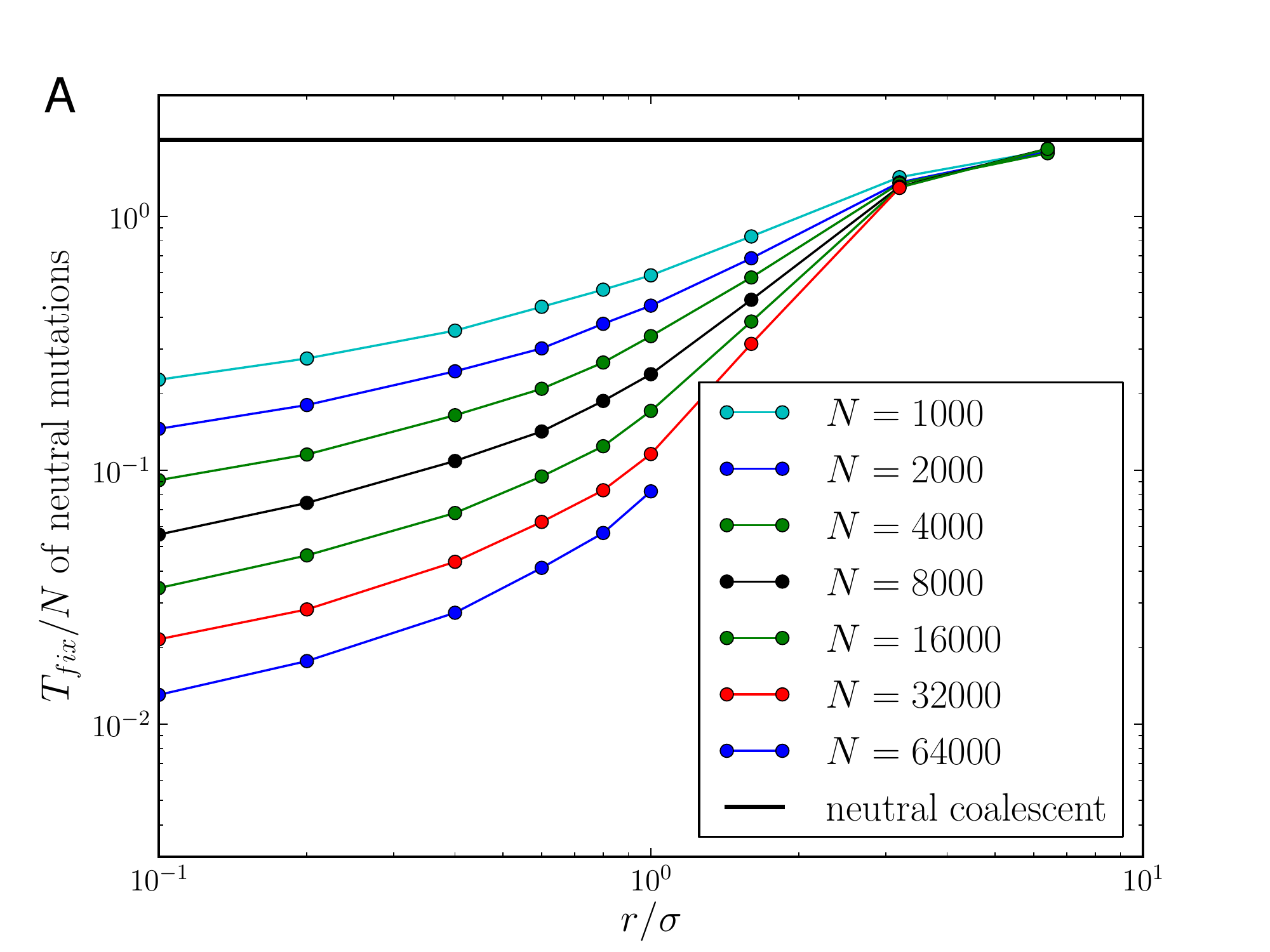}
  \includegraphics[width=0.45\columnwidth]{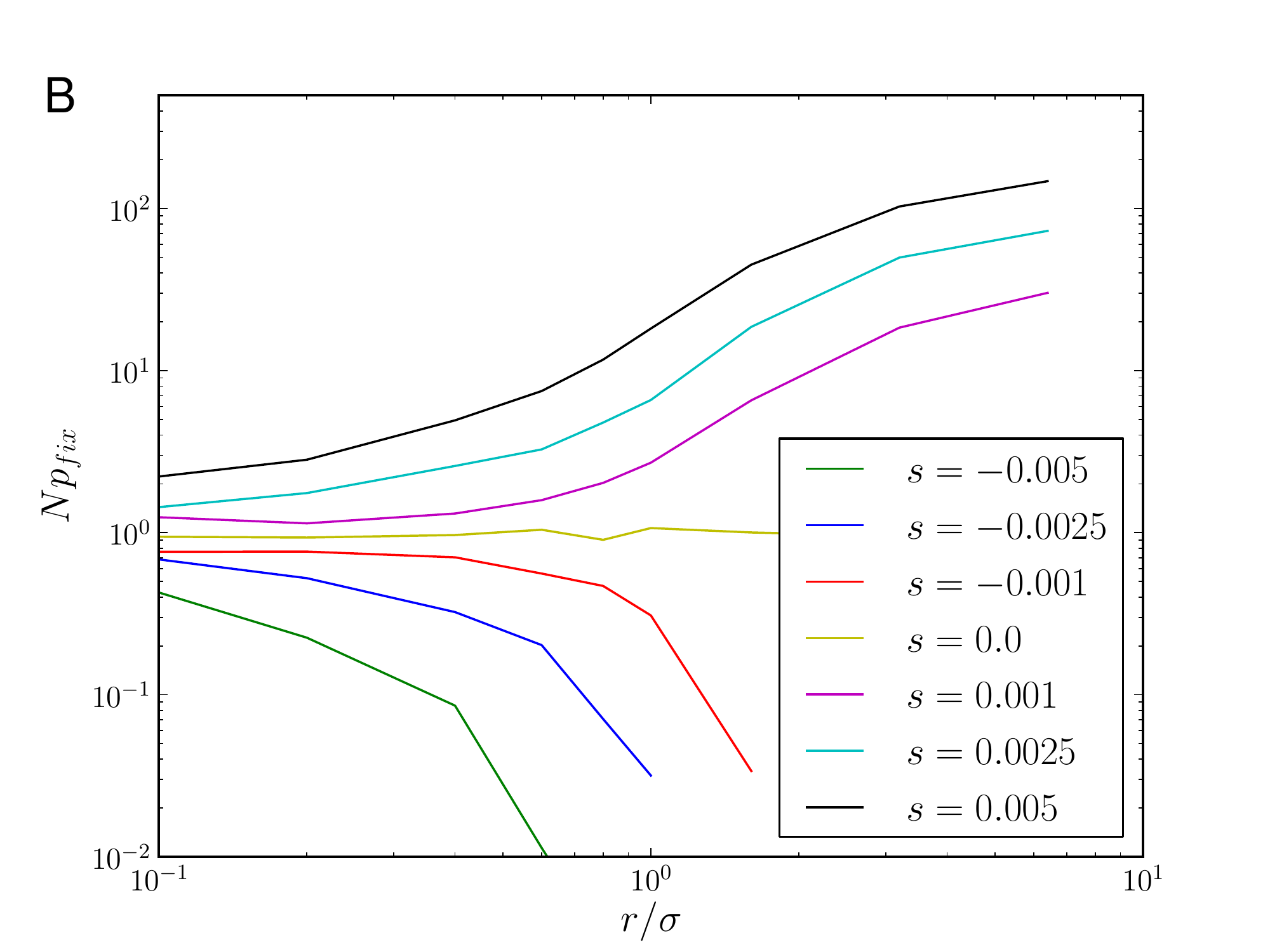}
  \caption[labelInTOC]{Genetic draft dramatically reduces fixation times and the efficacy of selection. Panel A) shows mean fixation times $\Tfix$ of neutral mutations, normalized by the populations size $N$, as a function of $\rec/\sigma$ for different $N$. For $\rec/\sigma < 1$, $\Tfix$ increases only slowly (sublinear) with $N$. Panel B) shows the fixation probability of beneficial and deleterious mutations relative to that of neutral mutations. Even though $|Ns|>50$, fixation probabilities are close to $N^{-1}$ at low outcrossing rates. ($N=16000$)}
  \label{fig:sim_Tfix_Pfix}
\end{center}
\end{figure}

\begin{figure}[htp]
\begin{center}
  \includegraphics[width=0.45\columnwidth]{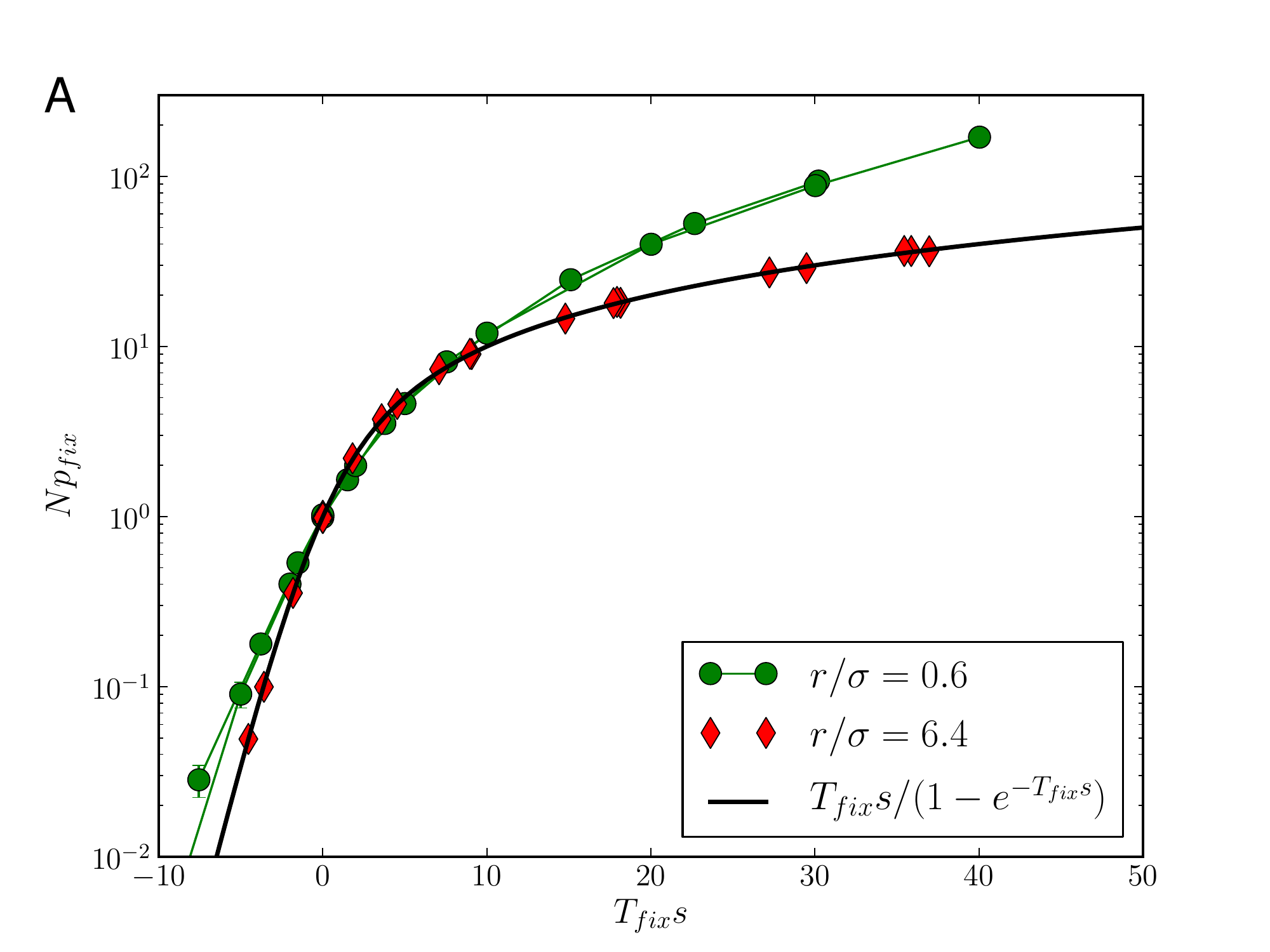}
  \includegraphics[width=0.45\columnwidth]{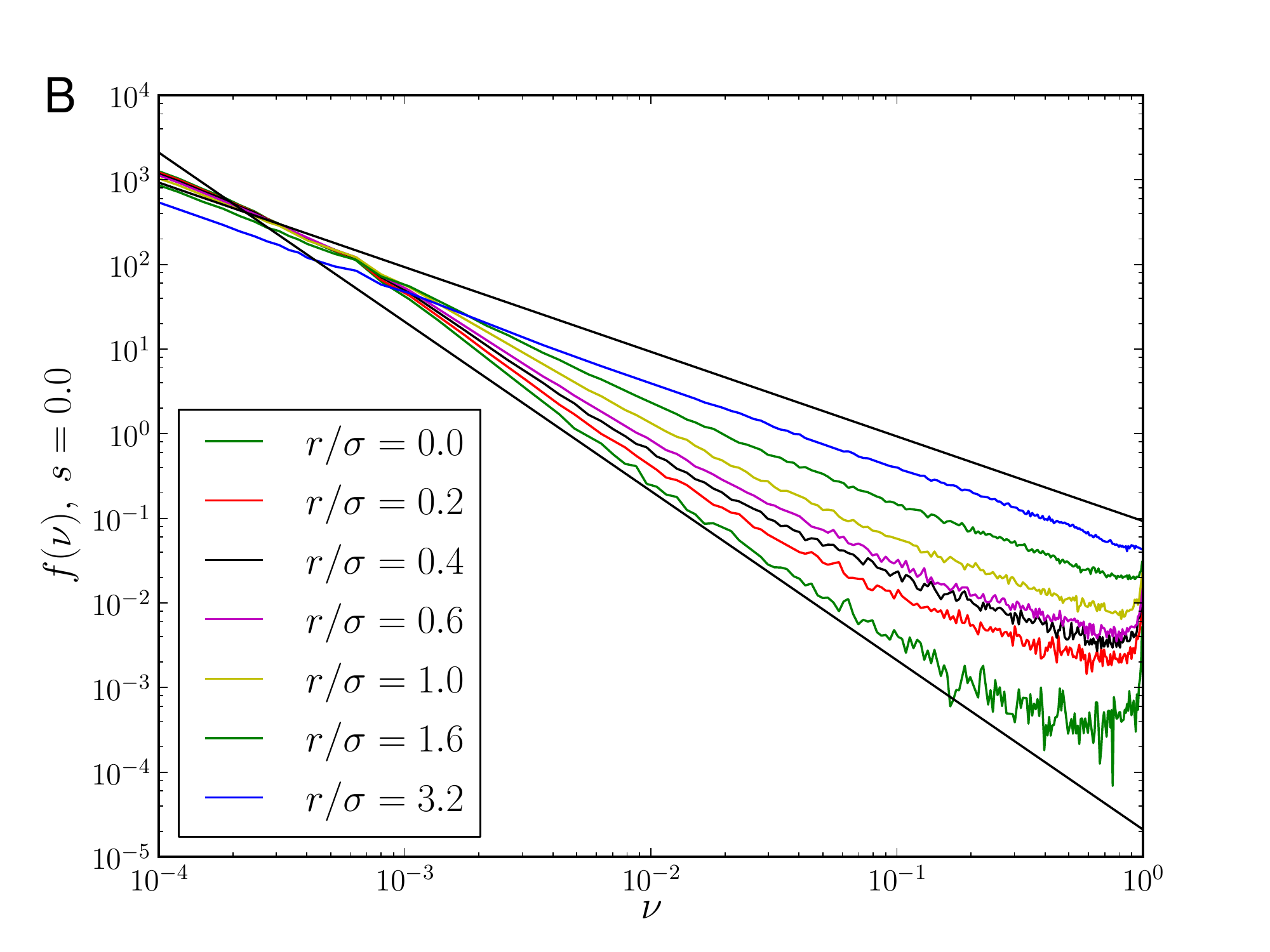}
  \caption[labelInTOC]{Fixation probabilities and allele frequency spectra are inconsistent with a neutral model with reduced effective population size. Panel A) shows the rescaled fixation probability $N\pfix$ of mutations with effect size $s$ as a function of $\Tfix s$, where $\Tfix$ is the fixation time of neutral mutations. In a neutral model, $\Tfix=2 N$, hence it seems sensible to define an effective populations size $\Ne=\Tfix/2$. The fixation probability of a mutation in a neutral model is indicated as black solid line. While the effective neutral model describes $\pfix$ well for $\rec/\sigma=6.4$, it fails for $\rec/\sigma=0.6$ (note the log scale). Panel B) shows the distribution of derived allele frequencies $\Pallele(\nu)$ in a population of $N=32000$ individuals for different $\rec/\sigma$. For $\rec/\sigma<1$, allele frequency spectra fall off much more steeply than the prediction $\Pallele(\nu)\sim \nu^{-1}$ of the neutral model. In fact, the distribution falls off more like $\Pallele(\nu)\sim \nu^{-2}$, which is the leading behavior predicted by Eq.~(\ref{eq:af_spectra}). All curves are normalized to unit area and the two power laws are indicated by straight lines.}
  \label{fig:sim_Pfix_collapse_afspec}
\end{center}
\end{figure}

Figure \ref{fig:sim_Tfix_Pfix} shows simulation results for the fixation time of neutral mutations and the fixation probability of beneficial and deleterious mutations for populations of different size and different outcrossing rates. At low $\rec/\sigma$, fixation times (Panel A) are reduced by orders of magnitude below the neutral expectation of $2N$ and reach the latter only for $\rec/\sigma$ is well above one, an effect already observed in \citet{Charlesworth:1992p40713}. The scaling of the data in panel A reveals that fixation times are not proportional to the population size at low $\rec/\sigma$, in which case the curves would lie on top of each other. Genetic draft has an equally dramatic effect on the efficacy of selection, which is shown in Fig.~\ref{fig:sim_Tfix_Pfix}b. The fixation probabilities of beneficial and deleterious mutations are only slightly perturbed from $N^{-1}$ if $\rec/\sigma<1$, even though $|Ns|$ ranges up to $80$. The fact that genetic draft reduces fixation times and the efficacy of selection is of course well known and it is customary to describe this effect by a neutral model with a reduced population size, i.e.~treating draft as if it were just an enhancement of genetic drift.  This, however, is often not appropriate, as draft causes fluctuations of a very different nature from that of genetic drift. To illustrate the problem, we define $2\Ne$ as the fixation time of neutral mutations $\Tfix$. With this definition, a mutation with effect size $\ds$ should fix with probability $\frac{\ds\Tfix}{N(1-e^{-\ds \Tfix })}$. However, the observed fixation probability does not follow this expectation if $\rec\leq \sigma$, as shown in Fig.~\ref{fig:sim_Pfix_collapse_afspec}a (note the logarithmic scale). 

Another striking feature where a neutral model with a reduced effective population size fails to capture the effects of draft is the allele frequency spectrum, shown in Fig.~\ref{fig:sim_Pfix_collapse_afspec}b for various ratios of $\rec/\sigma$. For small $\rec/\sigma$, the frequency spectrum $\Pallele(\nu)$ of neutral alleles falls off much more rapidly than the prediction of the diffusion theory: it decays as $\nu^{-2}$ with frequency instead of $\nu^{-1}$. This effect is often referred to as ``excess of rare variants'' \cite{Braverman:1995p34932}, but would, in fact, more aptly be called a ``lack of common variants''. Very similar dependencies of the fixation time, fixation probabilities, and allele frequency spectra on $\rec/\sigma$ are observed if fitness variation is not due to multiple selective sweeps but due to purifying selection. Simulation results for such a background selection scenario are shown in the supplementary figures S2 and S3.

To rationalize the observed behavior we solved analytically a simplified branching process model describing the dynamics of alleles in an adapting population; compare \emph{Mathematical model}. This branching process solution shows explicitly how observables depend on population parameters and elucidates the difference between draft and drift.

\subsection*{Analytic results for the branching process model}
\begin{figure}[htp]
\begin{center}
 \includegraphics[width=0.3\columnwidth]{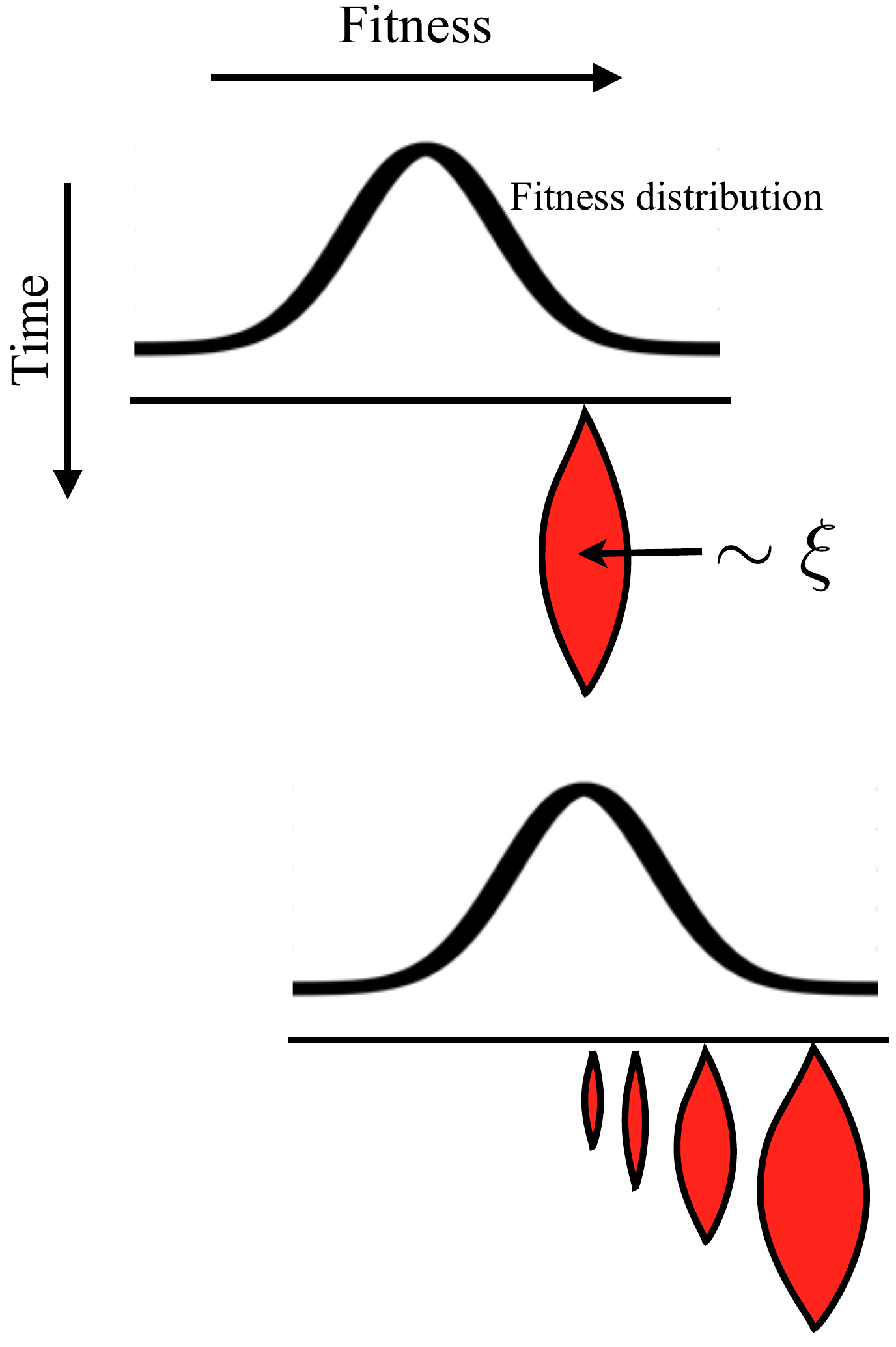}
 \includegraphics[width=0.65\columnwidth]{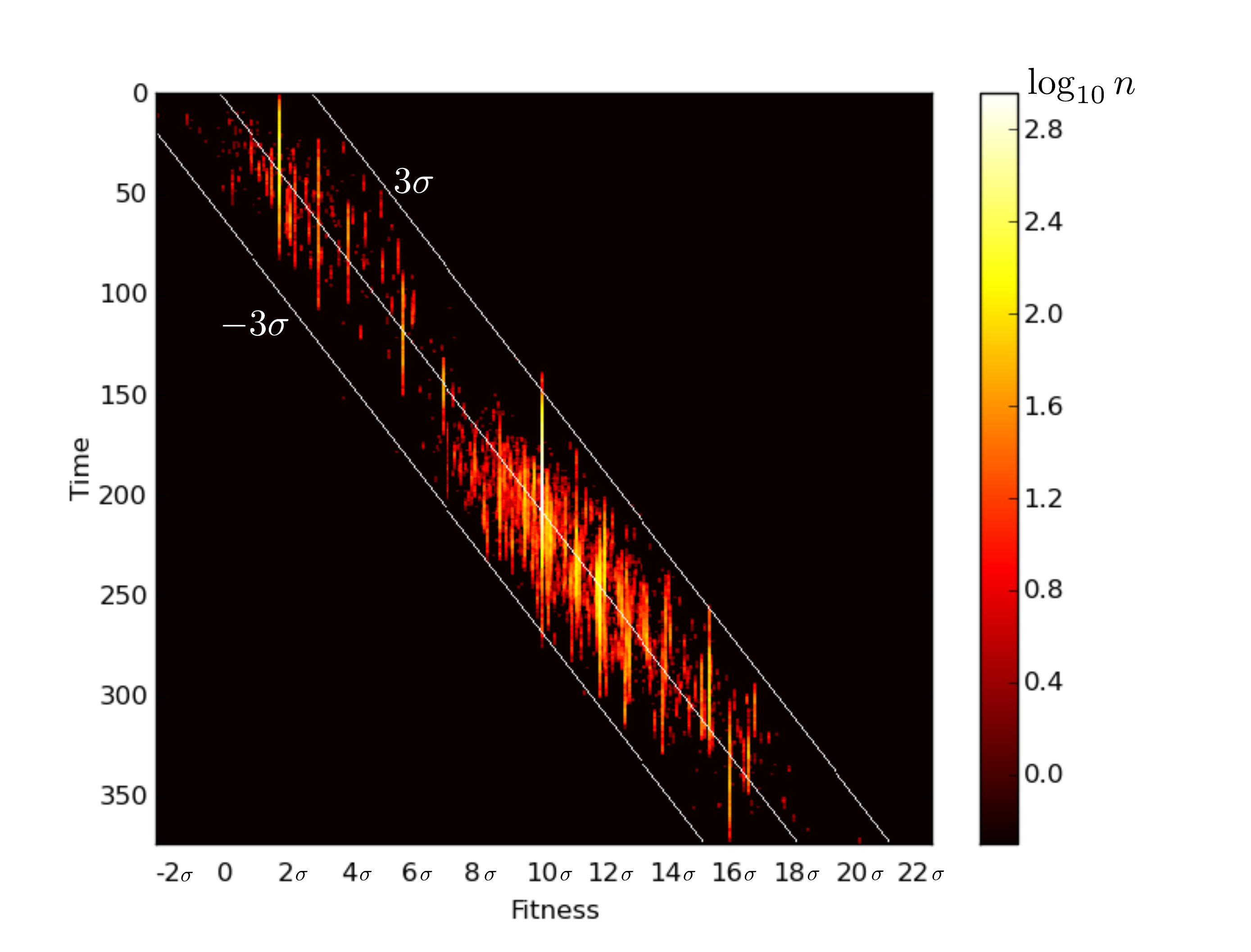}
  \caption[labelInTOC]{Clonal expansion results in large fluctuations. Panel A: A clone with high initial fitness can grow to large numbers before going extinct.  Large clones give rise to many recombinant daughter clones.  Panel B shows a corresponding trajectory obtained by simulation of the branching process equation, see \emph{Model}.  The mutant allele originated on a background with fitness $\x-\mfit(t)=2\sigma$ at time $t=0$. Each subclone of the population appears as a vertical line whose horizontal position indicates its fitness $\x$. The size of a clone (color coded) rises as long as $\x-\mfit(t)-\rec>0$ and shrinks after the population wave has passed. The larger a subclone, the more daughter clones it produces. The mean fitness $\mfit(t)$ as well as $\mfit(t)\pm 3\sigma$ are indicated as white lines. }
  \label{fig:illustration}
\end{center}
\end{figure}

The key to understanding the qualitatively different behavior at low $\rec/\sigma$ compared to $\rec/\sigma\gg 1$ is the fact that genotypes with fitness $\x$ can expand clonally if $\x-\mfit -\rec  >  0$, i.e.~if their growth rate exceeds the rate at which they outcross. In contrast to the case of $\rec/\sigma \gg 1$, for $\rec/\sigma\leq 1$ the condition for clonal expansion is fulfilled for a substantial fraction of the genotypes in a finite population. A genotype establishes  with probability $\x-\rec$ (setting $\mfit(t_0)=0$) and is clonally amplified so that its copy number subsequently grows as 
\begin{equation}
\label{eq:bubbles}
n(T, \x) \approx \frac{1}{\x-\rec}e^{(\x-\rec)T-\sigma^2 T^2/2} \,
\end{equation}
where $T$ is the time since establishment \citep{Desai:2007p954}. The term $\sigma^2 T^2/2 = \int_0^T dt \mfit(t)$ accounts for the increasing mean fitness, $\mfit(t)=\sigma^2 T$, of the rest of the population, competing with the clone. After establishment, a fit genotype 
therefore gives rise to a clone whose size has a Gaussian profile in time, as illustrated in Fig.~\ref{fig:illustration}a.
During its lifetime, an individual clone spawns an average of $\xi=\rec \int_0^\infty dt\, n(t)$ recombinant genotypes. For $\x-\rec\gg \sigma$ one finds $\xi \approx \sigma(\x-\rec)^{-1} e^{\frac{(\x-\rec)^2}{2\sigma^2}}$, which increases rapidly with increasing $\x-\rec$. In the Discussion, we will develop this intuition into a simplified model of the genealogy of clones.
Fig.~\ref{fig:illustration}b shows the copy number of the mutant allele on different genetic backgrounds as a function of time in the continuous time branching process model. The observed stochastic trajectory shows the features embodied in the simplified effective model: An anomalously fit genotype gives rise to many recombinant offspring genotypes, resulting in large fluctuations in the copy number of the mutant allele. 

The dynamics of the distribution $\Pnumber(n,T|\x)$ of a mutant allele, given that it arose on a genotype with fitness $\x$ ($k=1$, $t=0$), can be described by the branching process defined in Eq.~(\ref{eq:BME}).
We will proceed by analyzing the fixation probability, the typical time to fixation, and the allele frequency spectra for neutral, beneficial and deleterious mutations for the case $\rec/\sigma \leq 1$. In the opposite limit of rapid outcrossing, the effect of selection during the life time of a genotype is small and Eq.~\ref{eq:BME} can be analyzed perturbatively in $\sigma/\rec$. This perturbation analysis was performed in \citep{Neher:2010p30641}. It was shown that the fixation probabilities of beneficial mutations are reduced by a factor $1-\alpha \sigma^2/\rec^2$, where $\alpha$ is numerical factor of order one and depends on the details of the recombination model. This is consistent with the classical argument of \citet{Robertson:1961p36142} showing that the cumulative effect of draft is proportional to the square of the degree of linkage \citep{Santiago:1998p34629,Barton:2009p31030}.
The analysis and derivation of the results is mainly relegated to the appendix: here we present and discuss the results.

\subsubsection*{Extinction and survival of neutral mutations}
\begin{figure}[tp]
\begin{center}
  \includegraphics[width=0.48\columnwidth]{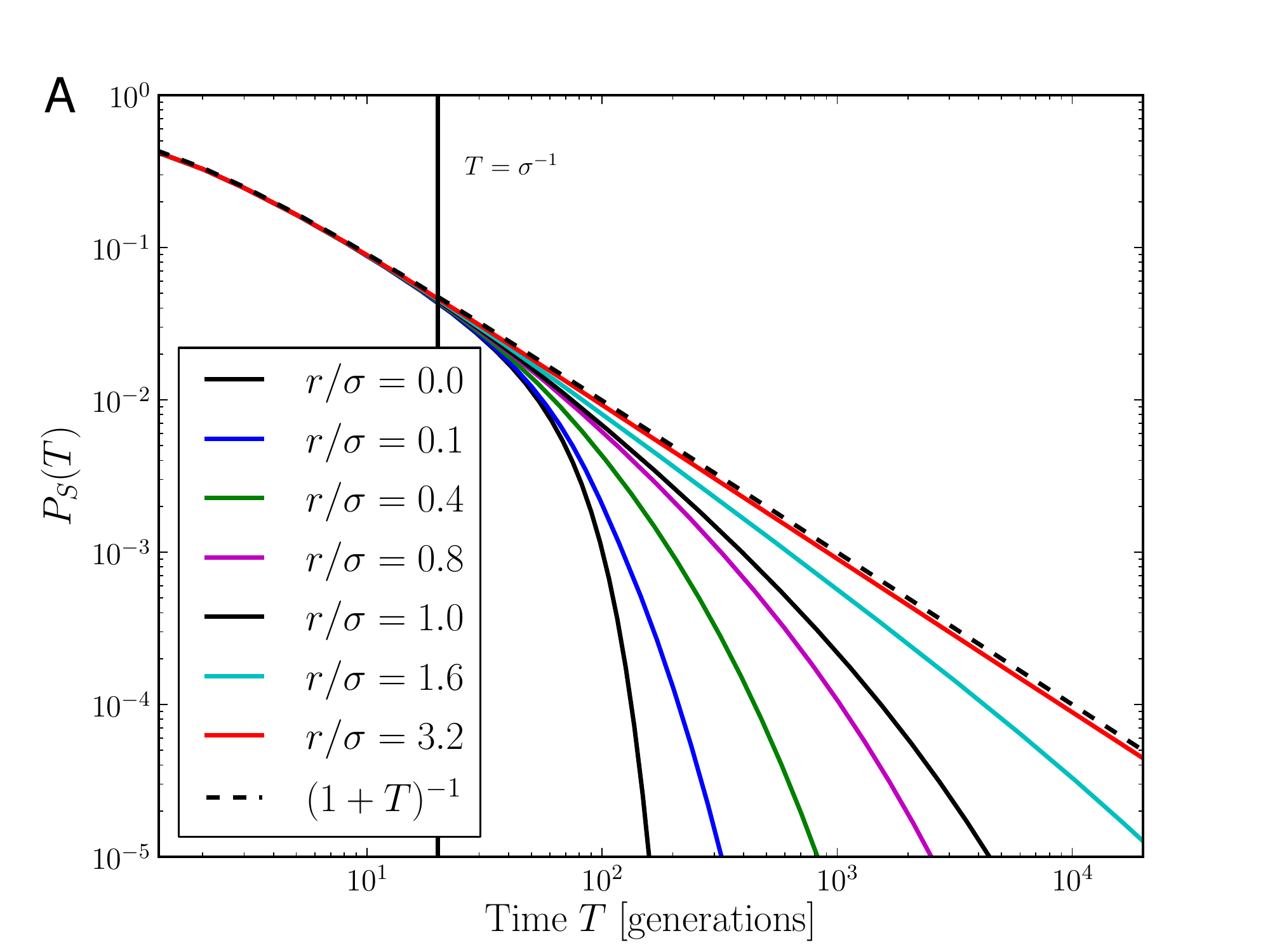}
  \includegraphics[width=0.48\columnwidth]{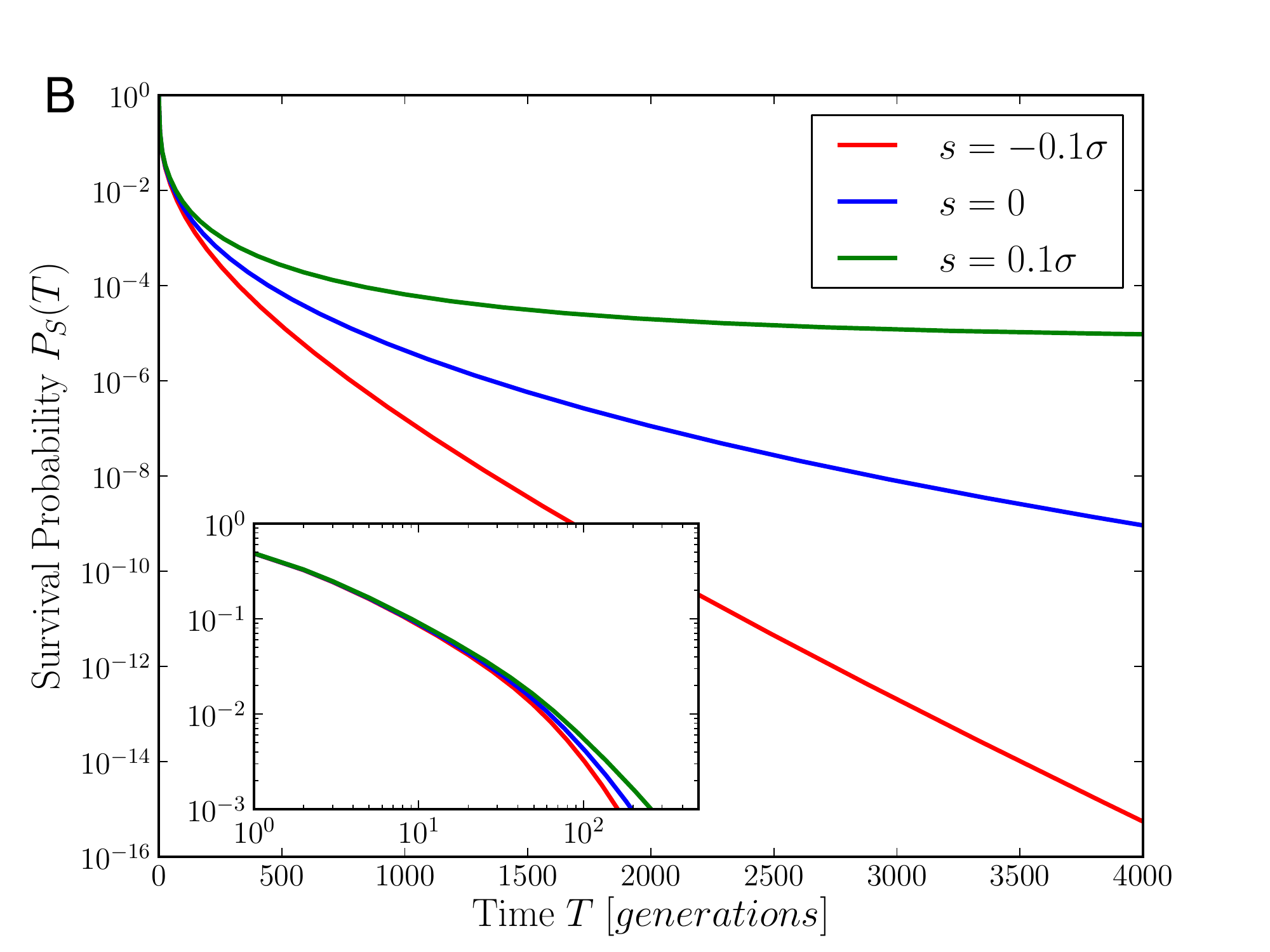}
  \caption[labelInTOC]{Survival of novel mutations in the branching process model. Panel A shows the survival probabilities of neutral mutations in the communal recombination model for different outcrossing rates as a function of time (obtained through numerical solution for the generating function of the branching process). For large $\rec/\sigma$, the survival probability closely follows the result for a single neutral locus $\Pnx = (1+T)^{-1}$ (note the log-log scale). For $\rec/\sigma \leq 1$, $\Pnx(T)$ decays much more rapidly as soon as $T\gg \sigma^{-1}$. Panel B illustrates how an intrinsic fitness effect $\ds$ affects the survival probability for $\rec/\sigma=0.4$ (note the log-linear scale). At short times, beneficial and deleterious mutations behave like neutral mutations. After a crossover time, which depends on $\ds$, the survival probability of deleterious mutations continues to decay exponentially with rate $\ds$, while the survival probability of beneficial mutations becomes time independent. $\Pnx(\tau)$ of neutral mutations ($\dss=0$) decays to zero with a decreasing rate.
  }   
  \label{fig:pool}
   \label{fig:quasineutrality}
\end{center}
\end{figure}

The fundamental quantity describing the stochastic dynamics of an allele is the distribution, $\Pnumber(n,T)$, of the number of copies that exist $T$ generations after the founding mutation. Note that the mutation could have arisen on genotypes with different background fitness $\x$, and $\Pnumber(n,T)$ is the average over all possible $\x$ at time $T=0$. Of particular importance is the survival probability $\Pnx(T)=1-\Pnumber(0,T)$. Fig. \ref{fig:pool} shows $\Pnx(T)$ for a neutral mutation for different outcrossing rates in the communal recombination model, obtained by a numerical solution of Eq.~(\ref{eq:phi}). Initially, $\Pnx(T)$ decays as $(1+T)^{-1}$, regardless of the outcrossing rate. This is the same behavior seen in homogenous or completely neutral populations. Selection on the heterogenous backgrounds starts influencing the fate of the allele at times $T>\sigma^{-1}$, i.e.~the reciprocal of the typical selection differentials in the population. After this initial transient, the survival probability depends strongly on the relative magnitude of the outcrossing rate and the fitness variance in the population, decaying faster for smaller $\rec/\sigma$.
We show in the Appendix (Eq.~(\ref{eq:longtime}) and Appendix \ref{sec:app_neutral}), that the time dependence of the survival probability is given by 
\begin{equation}
\label{eq:Pnx}
\Pnx(T) \sim \begin{cases}
        \sigma e^{-\frac{\sigma^2}{2\rec^2} \log^2 (\rec^3\sigma^{-2} T)} &  \sigma/\sqrt{\log N}<\rec <\sigma \\
	(1+T)^{-1} & \rec \gg \sigma
       \end{cases}
\end{equation}
Consistent with our previous argument, we find that for $\rec \gg \sigma$ neutral mutations are unaffected by selection on the background, reproducing the familiar result $\Pnx(T)\sim T^{-1}$. However, a strong deviation from this neutral behavior occurs as soon as $\rec/\sigma$ approaches one. In the regime with $\rec<\sigma$, the ratio $\rec/\sigma$ enters the dynamics of $\Pnx(T)$ as a prefactor of $\log^2 T$ in the exponent, resulting in faster extinction at lower $\rec$. (Yet another regime appears at even smaller $\rec$  for $\rec/\sigma < 1/\sqrt{\log N}$ and is discussed in the Supplement). 
Eq.~(\ref{eq:Pnx}) was derived using the communal recombination model. In Fig.~\ref{fig:Pnx_asymptotic} we show, via simulation of the branching process, that the infinitesimal model exhibits the same asymptotic dependence on parameters.

\subsubsection*{Dynamics of surviving alleles and fixation times}
Given that a new allele is not lost after $T$ generations, what is the mean number of its copies? The answer to this question is surprisingly simple and will inform us about the scaling of fixation times of neutral alleles. The unconditional expected number of copies after time $T$ is obtained by multiplying Eq.~\ref{eq:BME} by $n$, summing over $n$ and averaging over the distribution of $\x$  at $T=0$. One finds the simple expression
\begin{equation}
\overline{\langle n(T)\rangle} = e^{sT} ,
\end{equation}
which reduces to $\overline{\langle n(T)\rangle}=1$ in the neutral case. Here, $\langle\ldots\rangle$ denotes the average over $\Pnumber(n,T)$, while the overbar denotes the average over the fitness of the founding genotype.
Hence, the expected number of copies depends neither on recombination, nor on the speed of adaptation. It only depends on the intrinsic fitness effect of the novel allele. However, $\overline{\langle n(T)\rangle}$ is a product of the survival probability, times the average $n$ conditioned on survival. For neutral mutations, the mean $n$ conditioned on survival is therefore the reciprocal of the survival probability, and when $\Pnx(T)$ is of order $N^{-1}$, a non-extinct allele has reached copy numbers of $\sim N$ and has essentially fixed. The first passage time, $\Tfp$, to allele frequency $\nu$, i.e.~to $\nu N$ copies, is given asymptotically by 
\begin{equation}
\label{eq:Tfix}
\Tfp \sim \begin{cases} 
        \sigma^{2}\rec^{-3} e^{\rec\sigma^{-1} \sqrt{2\log \sigma\nu N}}  &  \sigma/\sqrt{\log N}<\rec <\sigma \\
		\nu N & \rec\gg \sigma  
           \end{cases}
\end{equation}
We see that the time to reach allele frequency $\nu$ scales very differently with $N$ in the two regimes.  Instead of diffusive dynamics with diffusion constant $\sim N^{-1}$, the rapid turnover of individuals by selection at $\rec<\sigma$ results in large fluctuations, which can propel a mutation rapidly to large numbers. Simulation results confirming the validity of the argument that inverse survival probability can be used as a proxy for typical copy numbers, as well as the adequacy of modeling allele frequency dynamics by a branching process are shown in Fig.~\ref{fig:Pnx_asymptotic}.

A different approach to calculate fixation times in a model of continuous adaptation with facultative out-crossing has been developed in  \citet{Rouzine:2007p17401,Rouzine:2010p33121}. \citeauthor{Rouzine:2010p33121} use a  coalescence approach based on the clonal structure of the population to calculate fixation times and obtain the same exponential dependence on $\rec\sigma^{-1}\sqrt{2\log N\rec}$.

\subsubsection*{Allele frequency spectra}
In addition to changing the dependence of the fixation time on the population parameters, draft also results in a qualitatively different allele frequency spectrum. This has been noted previously \citep{Braverman:1995p34932,Fay:2000p35077,Gillespie:2000p28513,McVean:2000p19278} but to our knowledge never calculated explicitly. Figure \ref{fig:sim_Pfix_collapse_afspec}b shows the distribution of derived neutral alleles frequencies $\nu=n/N$ at different outcrossing rates measured in individual-based computer simulations. At high outcrossing rates, we observe $\Pallele(\nu)\sim \nu^{-1}$ as expected for mutations subject to random drift alone. However, when draft is dominant, $\Pallele(\nu)$ decays more steeply as
\begin{equation}
\label{eq:af_spectra}
\Pallele(\nu)\sim \frac{e^{\rec\sigma^{-1}\sqrt{2\log N\nu\sigma}}}{\nu^2\log N\nu\sigma} \ .
\end{equation}
This steep decay persists up to a crossover frequency that depends on $\rec/\sigma$ and the population size. Beyond this crossover frequency, the finite population size constraint becomes important. The calculation of the allele frequency spectra is detailed in Appendix \ref{sec:app_copy_number} and the result of the copy number distribution is given in Eq.~(\ref{eq:Pn_asymptotics}).

The $\Pallele(\nu)\sim \nu^{-2}$ behavior can be understood heuristically by considering the sizes of clones containing the allele. Let's consider the asexual case with $\rec=0$.  To calculate the copy number distribution of a novel allele, $T$ generations after it originated, we have to average over the initial fitness $\x$ of the genotype that it could have arisen on. Given that the allele arose on a genetic background with fitness $\x$, its average number after $T$ generations is given by Eq.~(\ref{eq:bubbles}). Since we know that the distribution of $\x$ is $P(\x,0) = (2\pi\sigma^2)^{-1/2} e^{-\x^2/2\sigma^2}$, we can calculate the expected distribution of $n$ after time $T$ within the deterministic approximation of Eq.~(\ref{eq:bubbles}). One finds that $\Pnumber(n,T)\sim n^{-3/2}\exp(-\sigma^2 T^2/8-\frac{ \log^2 n}{2\sigma^2T^2})$. The allele frequency spectrum is an average over alleles of different age, $T$, and the above expression therefore has to be integrated over $T$, which yields $\Pallele(n)\sim n^{-2}$. In a facultatively sexual population with $\rec<\sigma$,  allele frequency spectra are also driven by the clonal expansion of individual genotypes. However, in contrast to the asexual case, additional clones are constantly seeded through outcrossing so that the novel allele resides in many different clones of different ages, similar to the average over asexual clones of different ages. For this reason, both the copy number spectrum of individual alleles as well as the frequency spectrum averaged over alleles of different ages have a leading behavior $\Pnumber(n,T)\sim n^{-2}$ and $\Pallele(\nu)\sim \nu^{-2}$. This reasoning is confirmed by the branching process calculation given in the Appendix \ref{sec:app_neutral}.

In addition to a much steeper spectrum at low frequencies, the spectra of neutral and deleterious mutations are non-monotonic and increase again at frequencies near one \citep{Fay:2000p35077}. On a linear chromosome with isolated sweeps that affect tightly linked neutral variants, this effect is easy to understand: A tightly linked variant is brought close to fixation, resulting in a pile-up of derived variants at frequencies close to one. In our case with many unlinked sweeps in a facultatively mating population the effect is less intuitive, but ultimately of similar origin. 

\subsubsection*{Selection efficiency and quasi-neutrality}
The rapid and erratic dynamics of alleles incurred by selection on other loci not only affects the time to fixation, but also the ability of selection to prune deleterious mutations and fix beneficial mutations. Mutations behave essentially neutrally, as long as selection on the intrinsic effect of the allele is outweighed by fluctuations. A beneficial mutation has established when it has risen to sufficient numbers that future extinction is improbable and the allele goes to fixation deterministically. Without draft, a beneficial mutation is established when it has reached copy numbers of about $\ds^{-1}$. With additional fluctuations through draft, however, the allele is more likely to go extinct and has to rise to much higher numbers to establish. Similarly, draft can propel deleterious mutations to copy numbers they would never reach under drift alone. This reduction of the efficacy of selection is a hallmark of Hill-Robertson interference \citep{Hill:1966p21029} and was mainly studied in obligate sexuals \citep{Barton:1994p34628,McVean:2000p19278}. Here, we explore these effects in facultative outcrossers.

Since an allele that has survived longer is likely to be present in more copies, the influence of stochastic fluctuations and hence the rate at which the mutant allele goes extinct decreases over time. This is explicit in the time derivative of the log survival probability, which in the regime $\rec<\sigma$ is given by (see Appendix \ref{sec:app_bene_dele})
\begin{equation}
\label{eq:dtPnx}
\partial_T \log \Pnx(T) \approx \ds-\frac{\sigma^2\log (\rec^3\sigma^{-2} T)}{\rec^2 T} 
\end{equation}
The first term $\ds$ accounts for the deterministic bias due to the intrinsic effect of the allele, which causes accelerated extinction if $\ds<0$ and preferential survival if $\ds>0$, comp.~Fig.~\ref{fig:quasineutrality}b. The second term accounts for the stochastic forces on the allele, whose importance decreases with $T$. The time after which the allele's fate is dominated by selection on its intrinsic fitness can be roughly determined by equating  the two terms in Eq.~(\ref{eq:dtPnx}). In the regime of intermediate outcrossing rates $\rec <\sigma$, we find this crossover time to be $T^* \approx \sigma^2\log(\rec/\ds)/\rec^2\ds$. If fixation of a neutral mutation occurs before that time, selection has little effect and the fate of the mutation is dominated by fluctuations until fixation. Hence, we find a window of quasi-neutrality for mutations with selection coefficient smaller than $\ds_c$ 
\begin{equation}
\label{eq:quasineutrality}
|\ds| <\ds_c = \frac{\sigma^2\log(\rec/\ds)}{\rec^2 \Tfix} \sim \rec e^{-\rec\sigma^{-1} \sqrt{2\log \sigma N}} \ ,
\end{equation}
within which the fixation probability of a mutation is $p_{fix}\sim N^{-1}$. Beneficial mutations with effects larger then $\ds_c$ have a chance of fixation $\pfix = \frac{\log \rec/\ds}{\rec}e^{-\frac{\sigma^2\log^2 \rec/\ds}{2\rec^2}}$, as already found in \citep{Neher:2010p30641}. Propagation of deleterious mutations with $|\ds| <\ds_c$ is also dominated by fluctuations, resulting in allele frequency spectra similar to neutral mutations. Only for $|\ds| >\ds_c$ does the allele frequency spectrum $\Pallele(\nu)$ decay exponentially as expected for deleterious mutations. Fixation of deleterious mutations is exponentially suppressed with $\pfix\sim N^{-1}e^{-|\ds|\Tfix}$. Simulation results, qualitatively showing  this behavior, are presented in Fig.~\ref{fig:sim_Tfix_Pfix}b and \ref{fig:sim_Pfix_collapse_afspec}.

A related phenomenon is observed in the context of a linear chromosome in the limit where only one sweep affects a polymorphism at any given time. \citet{Barton:1994p34628} showed that a beneficial allele that is subject to many {\it sequential}, weakly linked interfering sweeps has little chance of fixation if its selection coefficient is below a critical value. The critical value reported by Barton is $\ds_c\sim \sigma^2/R$, where $\sigma^2/R$ is the fitness variance of sweeps per map length. In this approximation, the frequency of the focal allele is repeatedly reduced by a factor that depends on the degree of linkage and the strength of the interfering sweeps: this results in an effective reduction in the growth rate of the frequency of the focal allele to $\ds-\ds_c$.  In the case of numerous {\it concurrent} sweeps that we consider, the situation is different because the focal allele is spread over many different backgrounds subject to selection. The older the allele, the more likely it is to be spread over many backgrounds, attenuating the effect of interference. Hence we find an additional dependence of $\ds_c$ on the typical fixation time scale and the extent of recombination that occurs during this time.

\subsection*{Stochastic tunneling and complex adaptations}
\label{sec:results_tunneling}
\begin{figure}[htp]
\begin{center}
  \includegraphics[width=0.48\columnwidth]{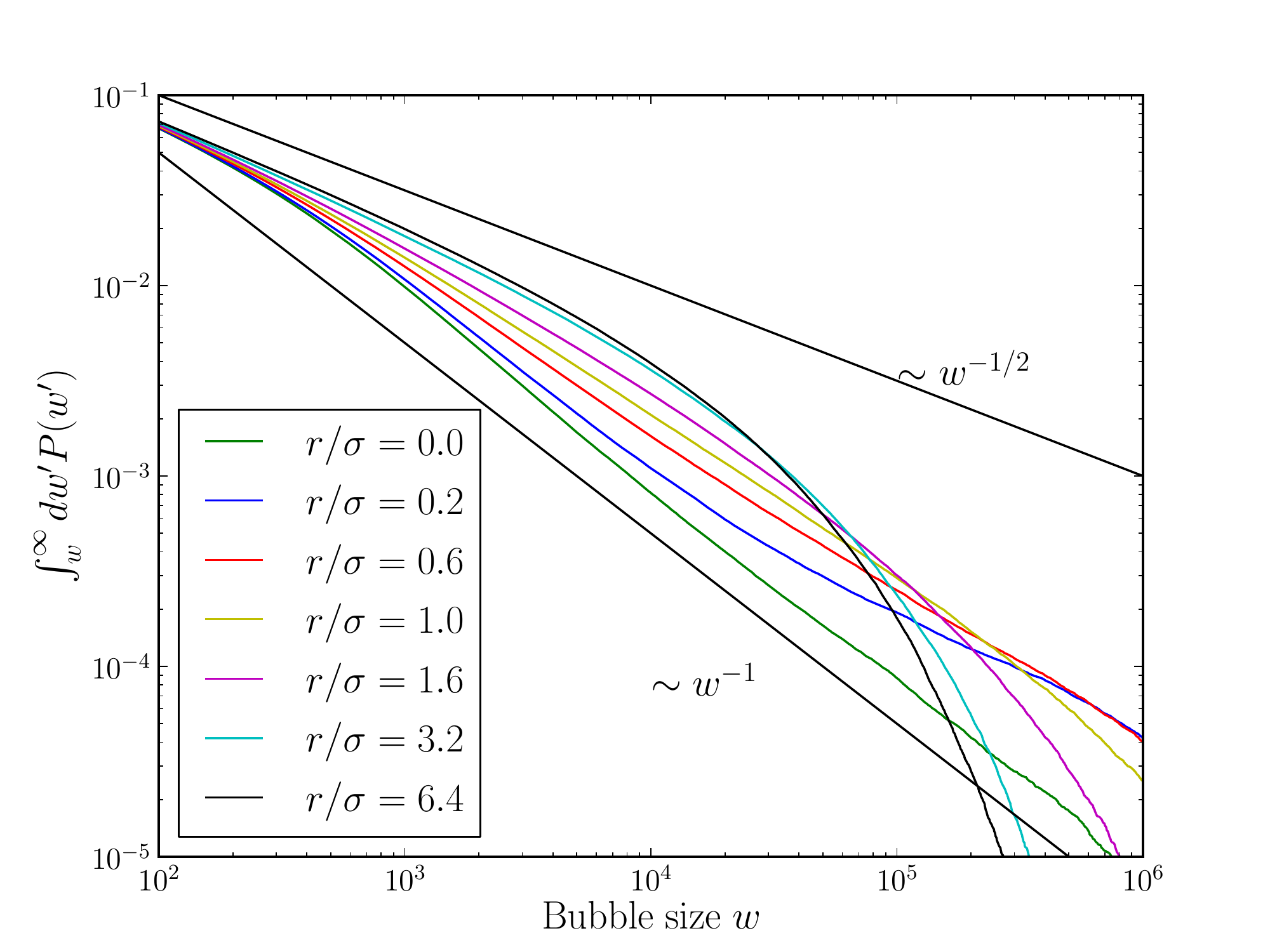}
  \caption[labelInTOC]{The cumulative distribution of ``bubble sizes'', i.e.~the integrated copy number $w=\int_0^\infty dt n(t)$, of deleterious ($\ds=-0.005$) alleles. The simulated bubble size distributions for $\rec/\sigma\ll 1$ agree with the predicted behavior $P(w)\sim w^{-2}$, which corresponds to a cumulative distribution $\sim \omega^{-1}$. For $\rec\gg \sigma$ the size of bubbles of deleterious alleles is cut off exponentially. While small bubbles are rarer at $\rec\ll\sigma$ compared to $\rec\gg \sigma$, this relation is reversed for large bubbles when the exponential cuts below the power law ($N=16000$). }
  \label{fig:bubbles}
\end{center}
\end{figure}

Consider an adaptation process where two mutations at the same locus are needed to confer a fitness advantage, but either mutation in isolation is neutral or deleterious. The rate of this process is given by the probability per unit time that the second mutation occurs in an individual already carrying the first mutation. More generally, the probability that the double mutant genotype never existed up to a time $T$ is 
\begin{equation}
P(T)=e^{-\mu \int_0^T dt (n_1(t)+n_2(t))} \ ,
\end{equation}
where $\mu$ is the mutation rate and $n_i(t)$ is the number of individuals carrying mutation $i$. The deleterious single mutants form a transient subpopulation in the large background population, to which we refer as a mutant ``bubble''. The important quantity determining the rate of such secondary events is the distribution of the cumulative number of single mutants $w(T)= \int_0^T dt n(t)$, i.e. the integrated bubble size. For homogeneous background populations, where genetic drift dominates the dynamics of neutral alleles, this quantity has been calculated by \citet{Iwasa:2004p32851,Weissman:2009p23398,Lynch:2010p36165,Weissman:2010p37077}.  In most cases of interest the rate of secondary events is small, so that the probability of tunneling in a single bubble is low. Hence tunneling takes longer than the lifetime of a typical bubble and only the long time limit of $w(T)$ is important. In analogy to the results presented above, we calculate the moment generating function $\Phi(z) = \langle e^{-zw}\rangle$ of the bubble size $w$, where $\langle \ldots\rangle$ denotes the average over the bubble size distribution $P(w)$. The calculation given in Appendix \ref{sec:app_tunneling} yields for the $\Phi(z)$
\begin{equation}
\Phi(z)=\begin{cases}
     ze^{\rec\sigma^{-1}\sqrt{-2\log z}} & |\ds| \ll \rec e^{-\rec\sigma^{-1}\sqrt{-2\log z}}\\
     \frac{z}{|\ds|} & |\ds| \gg \rec e^{-\rec\sigma^{-1}\sqrt{-2\log z}}
     \end{cases}
\end{equation}
The result for strongly deleterious mutations is the same as in homogenous populations \citep{Weissman:2009p23398}. The crossover between the two regimes occurs essentially at the quasi-neutrality threshold, with $N$ replaced by $z^{-1}$.  The generating function $\Phi(z)$, which is also the Laplace transform of $P(w)$, is of immediate practical relevance, since it is exactly the probability that a secondary event with rate $z$ does not occur within a single bubble. 
From the Laplace transform, we calculate the distribution of $P(w)$, which  in the draft dominated regime we find to be
\begin{equation}
P(w)\sim \frac{1}{w^2\sqrt{2\log w}}
\end{equation}
The latter has to be compared to the result for drift alone: $P(w)\sim w^{-3/2}$. Both of these behaviors are clearly seen in the simulation results shown in Fig.~\ref{fig:bubbles}. Draft causes large neutral bubbles to become rarer, but also dramatically enhances the probability of large bubbles for deleterious (quasi-neutral) alleles.

\section*{Discussion}
The importance of hitch-hiking and draft as a source of fluctuations was first discussed by \citet{Smith:1974p34217} as a possible explanation for the apparent insensitivity of genetic variation on the census population size, dubbed the ``paradox of variation'' \citep{Lewontin1974}. Draft is expected to dominate over genetic drift in very large populations and hence is the factor determining the level of genetic diversity and the efficacy of selection \citep{Barton:1995p3540,Gillespie:2001p9636}. It was shown to have appreciable effects on genetic diversity in \beast{Drosophila} \citep{Sella:2009p26729}. Draft is expected to be even more pronounced in facultatively outcrossing organisms, where alleles stay associated with each other for longer periods of time. Such organisms include many plants, fungi, nematodes, and viruses, including the human pathogens influenza and HIV. 

We have studied draft in continuously adapting facultatively outcrossing populations using direct computer simulation of a model inspired by intra-patient evolution of HIV. We explain the simulation results by analyzing a simplified branching process model, which is amenable to analytic calculations that elucidate how fixation time, the efficiency of selection and allele frequency spectra depend on fundamental parameters of the population dynamics. The simplification that makes these analytic results possible is a ``large number" limit: Due to the large number of concurrent sweeps, the distribution of fitness in the population assumes a Gaussian form and translates towards higher fitness with constant velocity. If the outcrossing rate, $\rec$, is comparable to or smaller than typical fitness differences, $\sigma$, between individuals in the population, clonal expansion of genotypes results in a heavy-tailed distribution of recombinant offspring with dramatic effects. In particular, we find that alleles with selection coefficients smaller than $|\ds_c|=\sigma^2\log (\rec/\ds)/(\rec^2\Tfix)$ behave \emph{quasi-neutrally}: Their fate is dominated by fluctuations and is nearly independent of the intrinsic allele fitness $\ds$. Yet because selection acting on the allele is masked not by random drift, but by {\it draft} - transient associations with chromosomal ``backgrounds" with different fitness - the behavior is quantitatively different from true neutrality. The conventional dynamics dominated by genetic drift is realized only in the limit of no linkage or complete neutrality (including the background). To emphasize this distinction we refer to the draft dominated regime as {\it quasi-neutrality}.

In addition to a reduced efficacy of selection, draft reduces fixation and coalescence times, which instead of growing linearly with the population size, increases much more slowly. The probability that a given neutral locus is polymorphic with an allele at intermediate frequency is roughly given by the product of the neutral mutation rate, $\mu$, and the coalescence time, $T_{C}$. Hence we expect the neutral diversity $\Theta\sim \mu \sigma^{2}\rec^{-3} e^{\rec\sigma^{-1} \sqrt{2\log \sigma N}}$. Details of the expected heterozygosity, however, depend on the allele frequency spectrum. The latter is also the most informative quantity that can be evaluated from a static population snapshot. We have shown that in our model, allele frequency spectra decay  as $\nu^{-2}e^{\rec\sigma^{-1}\sqrt{2\log N\nu\sigma}}$, which is much steeper than $\nu^{-1}$ predicted by drift. As a consequence, in contrast to the classic neutral theory \citep{Ewens_2004} one should expect the number of polymorphic loci to increase almost linearly (rather than logarithmically) with the sample size as one samples the population deeper and deeper. 

\subsection*{Why draft is different from drift}
Draft and hitch-hiking is often accounted for by an effective neutral model with reduced population size $\Ne$ \citep{Hill:1966p21029,Felsenstein:1974p23937}. This is possible if draft does not change the offspring distribution qualitatively but only increases its variance. The Central Limit Theorem then guarantees the existence of a meaningful diffusion limit for the allele frequency dynamics with the diffusion constant that could, if one were so inclined, be interpreted as the inverse effective population size. This diffusion limit, however, is not possible in our case since the number of recombinant genotypes, $\xi$, that descend from a single clone has a very broad distribution with power-law tails and diverging variance. To see this let's consider a genotype seeded by recombination at fitness $\x\gg \sigma$ above the mean, which establishes with probability $\x-\rec$. After establishment, its copy number $n(t)$ initially grows almost deterministically, see Eq.~(\ref{eq:bubbles}). Over time, the mean fitness increases such that the growth rate eventually turns negative and the genotype goes extinct.  During its lifetime, the clone produces $\xi(\x) \approx \rec \int_0^{\infty} dt \ n(t)\sim e^{(\x-\rec)^2/2\sigma^2}$ recombinant offspring. Since the initial fitness $\x$ of the genotype is drawn from a Gaussian distribution $P(x) = e^{-x^2/2\sigma^2}/\sqrt{2\pi\sigma^2}$, the mean and second moment of the distribution of the number of recombinant offspring are given by $\langle \xi\rangle = \int d\x P(\x)\xi(\x) = 1$ and $\langle \xi^2\rangle = \int d\x P(\x)\xi^2(\x)$, respectively. While the mean exists, $\xi^2(x) \sim e^{(\x-\rec)^2/\sigma^2}$ increases so rapildy with $\x$ that the second moment and hence the variance is divergent. Of course, the finite population cuts off the offspring distribution and strictly speaking prevents the divergence of the variance. However, this cut-off is on the order of the population size and is irrelevant for the dynamics of rare alleles and the existence of a diffusion limit. Below this cut-off, one finds that the distribution of $\xi$ behaves asymptotically as
\begin{equation}
\label{eq:bubble_number_distribution}
P(\xi)\sim \frac{e^{-\rec\sigma^{-1}\sqrt{2\log \xi\rec\sigma^{-1}}}}{\xi^2} \ .
\end{equation}
The implications of this heavy-tailed distribution of genotypes that descend from a single clone are best understood by slight abstraction of the model. Assume for a moment a population dynamics where clones are seeded simultaneously and their recombinant offspring only start growing after the last clone of the previous round has died, as illustrated in Fig.~\ref{fig:illustration}a. This corresponds to an effective discrete generation scheme, only that one generation corresponds to the rise and fall of a genotype, rather than a single individual. Each clone spawns a Poisson distributed number of recombinant genotypes with mean $\xi$, where the $\xi$ are drawn from the heavy tailed distribution $P(\xi)$. The generating function of the distribution of number of recombinant offspring is found to be $\hat{P}(\lambda) \approx e^{-(1-\lambda)(1-e^{-\rec\sigma^{-1}\sqrt{-2\log (1-\lambda)}})}$. Starting from a single genotype, we calculate the distribution $P_n(m)$ of the number of genotypes carrying the mutant allele after $n$ effective generations, which has the generating function
\begin{equation}
\hat{P}_n(\lambda) = 1-\phi\circ\phi\cdots \phi(\lambda) = 1-\Phi_n(\lambda) \ , 
\end{equation} 
where $\phi(\lambda)=1-\hat{P}(\lambda)$ and $\circ$ denotes functional composition, i.e. $f\circ g(x) = f(g(x))$. From this result, we derive a difference equation for $\Phi_n(\lambda)$ 
\begin{equation}
\Phi_{n+1}(\lambda)-\Phi_{n}(\lambda) = - e^{-\rec\sigma^{-1}\sqrt{-2\log(\Phi_n(\lambda)\rec\sigma^{-1})}} \Phi_{n}(\lambda) \ ,
\end{equation}
where it was assumed that $\Phi_n(\lambda)$ is small. This difference equation is the discrete analog of Eq.~(\ref{eq:Phidot}), which is derived in the appendix for the continuous time model and from which most of our results follow. Thus the dynamics of mutations in a rapidly adapting population can be viewed as population genetics of genotypes, rather than of individuals, with a power-law tailed offspring distribution with diverging variance. This feature makes the description by an effective neutral Fisher-Wright model impossible, since no diffusion limit exists when increments have such long-tailed distributions. 
A similar effect occurs in Gillespie's pseudo-hitch-hiking model \citep{Gillespie:2000p28513}, where a hitch-hiking event can bring an allele to instantaneous fixation.
 
Coalescent processes with broad and skewed offspring distributions are an active field of research, see for example \citep{Mohle:2001p41279,Schweinsberg:2003p40932}. It is known, that broad offspring distributions can result in simultaneous mergers of multiple lineages and have dramatic effects on the time to the most recent common ancestor, which can increase sublinearly with the population size. Whether and how our interpretation of the adapting population in terms of a coarse grained coalescent corresponds to a known universality class of coalescent models remains to be shown. Coalescent models with broad offspring distributions have been applied to diversity data of pacific oyster \citep{Eldon:2006p36003} and our results suggest that they might apply to a larger class phenomena.

\subsection*{Adaptation vs purifying selection}
Discussions of Hill-Robertson interference typically focus either on the effect of linked beneficial or deleterious alleles: The former is referred to as ``hitch-hiking" \citep{Smith:1974p34217,Kaplan:1989p34931}, the later as ``background selection" \citep{Charlesworth:1993p36005,Nordborg:1996p18149}. Our approach addresses the effect of variation in genetic background fitness independent of its origin. Thus, while our analysis was set up in the context of continuous adaptation driven by many simultaneous selective sweeps, it is equally applicable to fitness variation dominated by deleterious mutations. The balance between deleterious mutations and selection in large populations gives rise to a steady Poisson distribution of fitness \citep{Haigh:1978p37141,Charlesworth:1993p36005}.  If the deleterious mutation rate $U_d$ is much larger than the selection coefficient $s_0$ of mutations, the Poisson distribution is very close to Gaussian, as in the case of the adapting populations considered here. While the origin of fitness variation is very different in these two cases, the effect on the fate of mutations is similar. In the adaptation scenario, a genotype carrying the mutation in questions stays at a certain point along the fitness axis while the mean fitness is increasing. When deleterious mutations dominate, the mean fitness is constant and set by a mutation selection balance, while the fitness of asexual offspring of a particular genotype decreases due to accumulating weakly deleterious mutations. Since the dynamics is determined by fitness relative to mean fitness, there is little difference between these scenarios from the point of view of the focal mutation. We have simulated scenarios where fitness variation is due to purifying selection and found that the stochastic dynamics of novel mutations in a purifying selection scenario is very similar to the case where fitness variation is due to sweeping mutations. The analogs of Figs.~\ref{fig:sim_Tfix_Pfix} and \ref{fig:sim_Pfix_collapse_afspec} for background selection are shown in the supplementary Figs.~S2 and S3.

In the asexual case, the most relevant quantity for background selection is the size of the least loaded class given by $Ne^{-U_d/|s_0|}$, from which all future individuals descend and within which the dynamics is essentially neutral \cite{Haigh:1978p37141,Charlesworth:1993p36005}. With recombination, beneficial and deleterious mutations can decouple from each other and individuals in the bulk of the distribution have a chance of contributing to the future population, albeit with a smaller and smaller chance as the distance from the fittest class increases. In the limit of rapid recombination, this effect can be described by a reduction in the effective population, which has been studied by \citet{Hudson:1995p18197,Nordborg:1996p18149}. Our results imply that background selection changes the population genetics more dramatically when the outcrossing rate is small compared to the standard deviation is fitness.

In any real-world scenario, there will be contributions to fitness variation from beneficial as well as deleterious mutations. In fact, in HIV the fitness variation due to deleterious mutations is expected to be substantial: With a mutation rate of roughly 0.2 per genome replication \citep{Mansky:1995p38971} and an average effect size of a mutation of $s_0=-0.01$, we have $\sigma =\sqrt{U_d|s_0|} \approx 0.04$. The outcrossing rate is of the same order of magnitude \citep{Neher:2010p32691}, so our results are expected to apply for deleterious mutations alone.

\subsection*{Sequential versus multiple sweeps}
To illustrate the applicability of isolated versus multiple sweep regimes of selection we compare sweep frequencies in \beast{Drosophila} and in HIV. Genetic divergence between different \beast{Drosophila} species suggests a rate of amino acid substitution of about one every two hundred generations (see \citep{Sella:2009p26729} for review). While these estimates come with a rather large statistical and methodological uncertainty, they nevertheless indicate that sweeps are frequent, but don't interfere. Each sweep ``occupies" a stretch of $\sim \ds/\rho$ nucleotides ($\rho$ is the recombination rate per nucleotide) on a chromosome for $\sim \ds^{-1}$ generations. Since they come in at a rate of 0.005 per generation spread over the total map length, they should on average be far apart: expect on average less than 1\% of genome length to be subject to draft at any given time. However, sites under weak selection might still cause substantial Hill-Robertson interference \citep{McVean:2000p19278}.

\begin{figure}[htp]
\begin{center}
  \includegraphics[width=0.48\columnwidth]{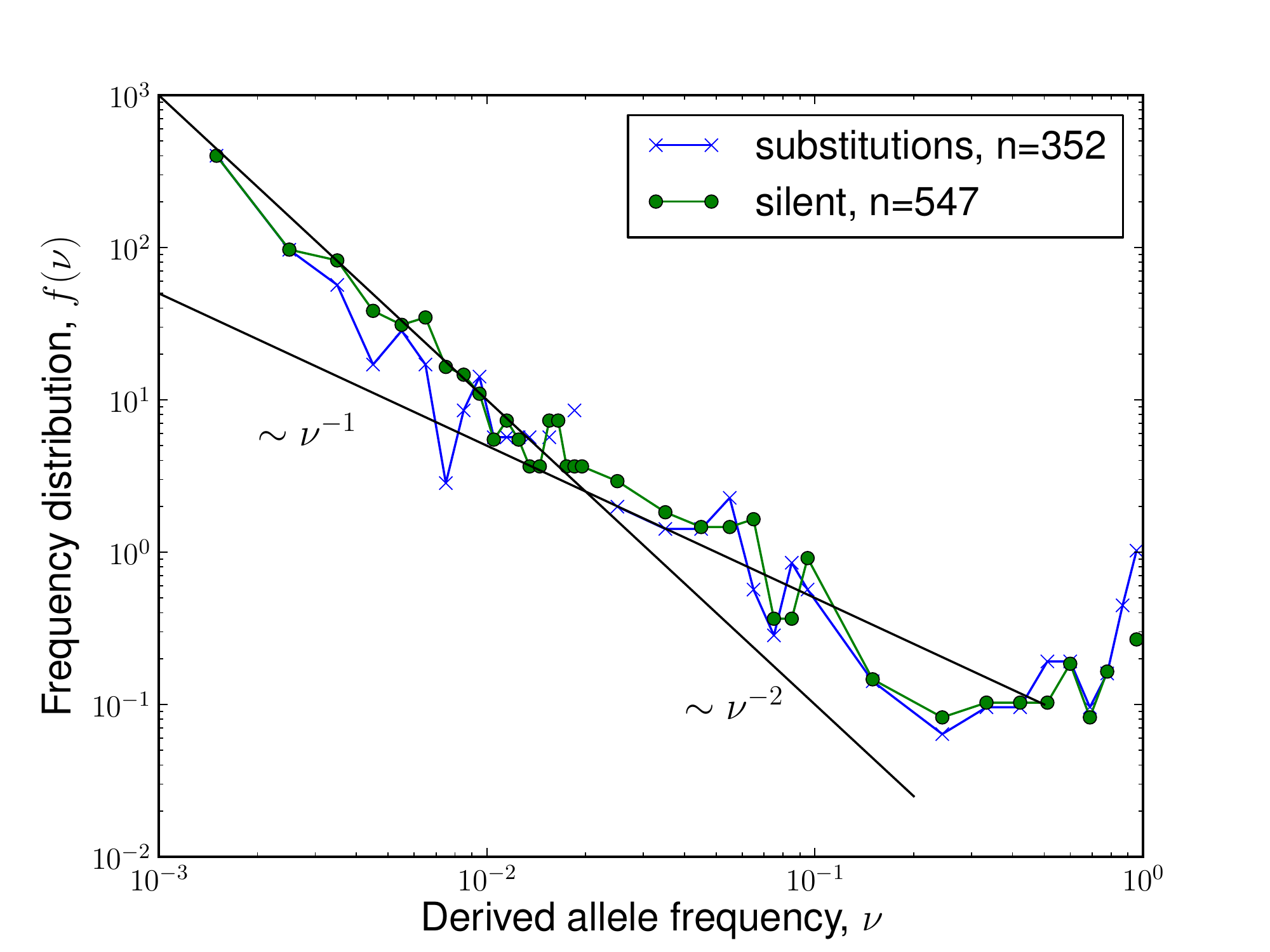}
  \caption[labelInTOC]{Derived allele frequency spectrum of silent and non-synonymous mutations in the HIV \gene{pol} gene, aminoacids 180-220 \citep{Hedskog}. The black lines indicates the expectation $\Pallele(\nu)\sim \nu^{-2}$ with draft and $\Pallele(\nu)\sim \nu^{-1}$ in absence of draft. The neutral spectrum $\sim\nu^{-1}$ fails to capture the steep decay at low frequencies and fits the data only in an intermediate frequency range. Note, however, that population expansions can give rise to allele frequency spectra similar to those expected under draft. }
  \label{fig:HIVsfp}
\end{center}
\end{figure}

On the other hand, the multiple sweep regime is likely to be relevant to organisms with a facultative sexual life cycle. In particular pathogens like HIV are under constant selection pressure resulting in a high rate of selective sweeps. Selection in HIV evolution is best characterized in the \gene{pol} gene, the target of  most anti-retroviral drugs, and in the \gene{env} gene, the target of neutralizing antibodies. On the order of 100 codons are implicated in drug resistance \citep{Chen:2004p22606,Rhee:2003p24151} and several such mutations compete and sweep during the evolution of drug resistance in a single patient. The \gene{env} gene frequently builds up a nucleotide diversity of 3-4\% \citep{Shankarappa:1999p20227}, with frequent adaptive amino-acid substitutions \citep{Williamson:2003p26136,Neher:2010p32691}. The effective recombination rate is about $10^{-5}$ per base and generation\citep{Levy:2004p23309,Neher:2010p32691}. With typical selection strength of a few percent per generation, the characteristic length of the segment affected by the sweep is about 1kb, which is comparable to the size of the evolving genes: hence the dynamics is in the multiple sweep regime. While \gene{pol} and \gene{env} genes constitute only a fraction of the HIV genome, cytotoxic T-lymphocyte (CTL) epitopes are found throughout the HIV genome \citep{LANLImmunology} so that even more sweeps associated with CTL escapes are expected \citep{Asquith:2006p28003}. In a recent study, \citet{Hedskog} characterized the evolution of the \gene{pol} gene of HIV in 6 longitudinally sampled patients during anti-retroviral treatment. Using their data, we measured the site frequency spectrum of derived alleles, shown in Figure \ref{fig:HIVsfp}. The allele frequency spectrum is indeed much steeper than expected for neutral evolution and compatible with $\nu^{-2}$, rather than the neutral expectation of $\nu^{-1}$. Furthermore, there is little difference between the spectrum of silent and non-synonymous mutations, consistent with our notion of quasi-neutrality. Note, however, that steep allele frequency spectra proportional to $\nu^{-2}$ are also expected in expanding populations.  While the sequences used in Figure \ref{fig:HIVsfp} are from chronically infected patients, changes in drug therapy resulted in shrinking and expanding population sizes and the effects of this expansion and draft cannot be disentangled at present.

In a recent study, \citet{Rouzine:2010p33121} present an analysis of adaptation of HIV using a model similar to the one used here and in \citep{Neher:2010p30641}.  Rouzine and Coffin study the problem of selection on standing genetic variation and in addition to the speed of adaptation, they compute the rate of coalescence, which in their model is controlled by the probability that two individuals in the population are genetically identical, i.e.~are part of the same clone. In agreement with our result Eq.~(\ref{eq:Tfix}), they find that the coalescence rate is $\sim e^{-\frac{r\x_{max}}{\sigma^{2}}}$, where $\x_{max}$ is the fitness of the fittest individuals in the population (relative to $\bar{\x}$). The quantity $\x_{max}$ corresponds to $\sigma\sqrt{2\log \sigma N}$ in our Eq.~(\ref{eq:Tfix}). Furthermore, \citet{Rouzine:2010p33121} study how benefical alleles at low frequency can be lost due to the lack of recombination. Our work complements \citet{Rouzine:2010p33121}, as we study the stochastic dynamics of new alleles arising from mutations, rather than the evolution of the fitness distribution, given standing variation. On the other hand, \citet{Rouzine:2010p33121} account for correlations between loci by allowing the distribution of recombinants to be broader than the population distribution \citep{Rouzine:2005p17398}- an effect absent in our model. In the range of recombination rates where our results apply, the latter deviations are small.

Our analytic results are derived using a branching process approximation that assumes that the novel allele is a small fraction of the total population and the finiteness of the population does not yet affect its dynamics. The influence of the finite population size is apparent in Figure \ref{fig:sim_Pfix_collapse_afspec}b, causing deviations from the branching process predictions at large frequencies. The branching process approach can be complemented by an effective theory for macroscopic frequencies, similar in spirit to diffusion theory for random drift. This theory, however, has to account for the very broad distribution of clone sizes. Throughout the manuscript, we have assumed that the speed of adaptation is identical to the variance in fitness, i.e.~we have assumed that effects on fitness of different mutations are purely additive. If interactions between mutations contribute to fitness, the variance in fitness tends to be larger than the speed of adaptation, since coadapted combinations are destroyed by recombination. Including genetic interactions within our framework  involves decoupling $v$ and $\sigma$, as well as changing the recombination functions. Genetic interaction might increase the importance of draft significantly, as interactions have a tendency to result in clonal population structures and couple the loci along the genome beyond physical linkage \citep{Neher:2009p22302}. Another simplification we have made is to assume that all loci segregate independently in an outcrossing event. Our analysis can be extended to account for linear chromosomes by considering a hierarchy of recombination rates for chromosomal segments at different distances. These, and other extension are interesting projects, which we leave for future work.

Several large-scale re-sequencing projects are underway with the goal to characterize genetic diversity in populations of humans, Drosophila and Arabidopsis at great depth. This upcoming data will allow us to measure allele frequency spectra to much greater depth, similar to the example from HIV shown in Figure~\ref{fig:HIVsfp}, and reveal the mechanisms shaping genetic diversity in natural populations.

\subsection*{Acknowledgements}
The authors acknowledge stimulating discussions with Daniel Fisher and Michael Lynch, thank Sally Otto and Nick Barton for a critical reading of the manuscript and thank Jan Albert for providing the HIV polymorphism data prior to publication. This research was supported by the National Science Foundation under Grant No. PHY05-51164 (RAN and BIS), NIH R01 GM-086793  (BIS) and the Harvey L.~Karp Discovery Award to RAN.

\appendix
\section{Appendix: Derivation of the results}
\label{sec:derivations}
To analyze the dynamics of the probability distribution, it is useful to
consider the generating function $\hat{p}(\lambda, T, t,\x) = \sum_n \lambda^n
\Pnumber(n,T|1,t,\x)$. From Eq.~\ref{eq:BME}, one obtains (see Supplement) the following equation for
$\phi(\lambda, T,t, \x) = 1 - \hat{p}(\lambda, T, t,\x)$:
\begin{equation}
\begin{split}
\label{eq:phi}
-\partial_t \phi(\lambda, T, t,\x)=
\rec\int d{\x'}K_{\x\x'}\phi(\lambda,T,
t,\x')+(\x-vt+s-\rec)\phi(\lambda, T, t,\x)-\phi^2(\lambda, T, t,\x)
\end{split}
\end{equation}
with boundary condition $\phi(\lambda, T, T,\x)=1-\lambda$. The survival probability is given by evaluating $\phi(\lambda, T, t,\x)$ at $\lambda=0$, while the moments and asymptotic behavior of $\Pnumber(n,T,t)$ are determined by the behavior of $\phi(\lambda, T, t,\x)$ in the vicinity of $\lambda=1$. We have to solve for $\phi(\lambda, T, t,\x)$ in these two limits, which will show up repeatedly below. It is convenient to remove the explicit dependence on the initial condition and the velocity $v=\sigma^2$ by rescaling $\psi(T,t,\x)= \phi(\lambda, T,t,\x)/(\sigma \epsilon)$ with $\epsilon=1-\lambda$. The two limits of interest are now $\epsilon\ll 1$ ($\lambda\approx 1$) and $\epsilon\approx\sigma^{-1}\gg 1$ ($\lambda\approx 0$). After rescaling $\rs = \sigma^{-1} \rec$, $\dss = \sigma^{-1} \ds$, and
$\xs=\sigma^{-1} \x-\sigma T$, and $\tau = \sigma (T-t)$, Eq.~\ref{eq:phi} takes the form 
\begin{equation}
\begin{split}
\label{eq:psi}
\partial_\tau \psi(\tau, \xs)=
\rs\int d\xs' K_{\xs\xs'}\psi(\tau, \xs')+(\xs+\tau+\dss-\rs)
\psi(\tau, \xs)-\epsilon \psi^2(\tau, \xs) \ ,
\end{split}
\end{equation}
where $\psi(\tau, \xs)$ is the rescaled generating function of the copy number distribution of a mutation that initially occurred on genome with rescaled background fitness $\xs$. The distribution of rescaled fitness $\xs$ in the population $P(\xs,\tau)=\frac{1}{\sqrt{2\pi}}e^{-(\xs-\bar{\xs}(\tau))^2/2}$ is Gaussian around the moving mean fitness $\bar{\xs}(\tau)$. Averaging Eq.~\ref{eq:psi} over $P(\xs,\tau)$ yields an equation for the scaled generating function $\Phi(\tau)=\int d\xs P(\xs,\tau)\psi(\tau, \xs)$.
\begin{equation}
\label{eq:solvability}
\partial_\tau
\Phi(\tau) =\partial_\tau \int d\xs P(\xs,\tau)\psi(\tau, \xs) =  \dss\Phi(\tau) - \epsilon\int d\xs P(\tau, \xs)\psi^2(\tau, \xs)
\end{equation}
For beneficial mutations, $\Phi(\tau)$ approaches a steady state where the two terms on the right balance each other. This long time limit was used as solvability condition in \citet{Neher:2010p30641} to determine the fixation probability of beneficial mutations. The steady state is independent of the initial condition $\epsilon$ and therefore equals the rescaled survival probability. However, if $\dss \leq 0$, no such steady state exists, since neutral and deleterious mutations have zero chance of fixation in an infinite population. To calculate the probability that a mutation reaches $n$ copies after time $\tau$, we need to find the full time dependent solution of Eq.~\ref{eq:solvability}. For simplicity, we will use the communal recombination model throughout and show that the infinitesimal model yields similar results by simulations of the latter. For the communal recombination model, the recombination contribution in Eq.~\ref{eq:psi} reduces to $\rs\Phi(\tau)$. 

In analogy to \citet{Neher:2010p30641}, we solve Eq.~\ref{eq:psi} in the two asymptotic regimes of large positive and large negative growth rate $\gr=\xs+\tau+\dss-\rs$.  For $\gr<0$ the non-linearity in Eq.~\ref{eq:psi} can be neglected since $\psi(\tau, \xs)\ll \epsilon^{-1}$. Conversely, at large $\gr$ the dominant balance in Eq.~\ref{eq:psi} is $\gr\psi(\tau, \xs)=\epsilon\psi^2(\tau, \xs)$. The two asymptotic solution are:
\begin{equation}
\label{eq:linear_solution}
\psi(\tau, \xs)=\begin{cases} 
             \rs e^{\gr^2/2}\int_0^\tau
d\tau' \Phi(\tau')e^{-\gr'^2/2} & \gr<\Thc \\
	\gr/\epsilon & \gr>\Thc
             \end{cases}
\end{equation}
The crossover between the two solutions occurs in narrow region at $\gr=\Thc (\tau)$ where $\Thc (\tau)$ is approximately the point where the two solutions cross. The evolution of $\Phi(\tau)$ is governed by Eq.~\ref{eq:solvability}, which has to be self-consisted with Eq.~\ref{eq:linear_solution}. We will first solve for $\Phi(\tau)$ assuming that $\tau$ is large. In this case the dominant contribution to the integral in Eq.~\ref{eq:linear_solution} will come from a well-defined $\tau'\gg 0$ which is determined by the maximum of $\exp(-\gr'^2/2+\log \Phi(\tau'))$ at $\tau' = \tau-\gr$. $\Phi(\tau)$ will turn out to change slowly and its time derivative can be neglected when determining the maximum of the exponent. Hence, 
\begin{equation}
\psi(\tau)\approx\sqrt{2\pi}
\rs\Phi(\tau)e^{\frac{\gr^2}{2}}
\end{equation}
This solution is valid below $\Thc(\tau)$, where $\psi(\tau)$ crosses over the
linear saturated form $\gr/\epsilon$. Substituting this solution into  
Eq.~\ref{eq:solvability} yields 
\begin{equation}
\label{eq:Phidot}
\partial_\tau\Phi(\tau) = \dss\Phi(\tau) - \rs\Phi(\tau)e^{-\rs\Thc(\tau)-\rs^2/2} \ ,
\end{equation}
We will first consider neutral mutations, $\dss=0$, in which case 
$\Phi(\tau)^{-1}\partial_\tau \Phi(\tau)=-\rs e^{-\rs\Thc(\tau)-\rs^2/2}$, confirming that $\log \Phi(\tau)$ changes slowly as long as $\rs\Thc\gg 1$ (the limit of $\rs\Thc \ll 1$ yields qualitatively different answers). Solving Eq.~(\ref{eq:Phidot}) for $\Thc(\tau)$ and differentiating with respect to $\tau$ yields an equation for $\partial_\tau \Thc(\tau)$, which is readily solved for $\Thc(\tau)$ in case $\rs\Thc\gg 1$
\begin{equation}
\Thc(\tau)=
\label{eq:longtime_theta}
\frac{1+W(\rs^3(\tau-\ti)e^{-1})}{\rs}\approx \frac{\log (\rs^3(\tau-\tau_0)+e^{\rs\Thi})}{\rs} 
\end{equation}
where $\Thi$ is the initial condition at $\tau=\ti$ and $W(x)$ is Lambert's $W$, i.e.~the solution of $W(x)e^{W(x)}=x$. The approximation of $W(x)$ by a logarithm is only accurate at very large time since there are significant $\log\log x$ corrections. The solution for the rescaled generating function then reads
\begin{equation}
\label{eq:longtime}
\Phi(\tau)= \frac{\Thc(\tau)}{\epsilon\rs}e^{-\frac{\Thc(\tau)^2}{2}} \sim
e^{-\frac{\log^2 (\rs^3(\tau-\ti)+e^{\rs\Thi})}{2\rs^2}}
\end{equation}
where the last terms only include the exponential factors. Apart from the rescaling, this solution is independent of the initial condition $\epsilon$. The dependence on $\epsilon$ will reemerge through the matching to the short time solution.  Note that $\Phi(\tau)$ satisfies asymptotically for large $\tau$:
\begin{equation}
\label{eq:non_extinction_rate}
\partial_\tau \Phi(\tau) \approx - {\log(\rs^3 \tau ) \over \rs^2 \tau } \Phi(\tau)
\end{equation}
which means that the relaxation rate of $\Phi$ decreases with time, but more slowly than it does in the case of pure drift (in the absence of draft), in which case $\partial_\tau \log \Phi(\tau) = - 1/\tau $.

The long time solution in Eq. ~\ref{eq:longtime} has to be matched to the solution at small $\tau$ where the initial condition cannot be neglected. 

\subsection{Short-time solution}
\label{sec:app_shortime_solution}
To analyze the short time behavior, it is useful to split $\psi(\tau, \xs)=\psi_r(\tau, \xs)+\psi_0(\tau, \xs)$ into
a contribution fed by recombination $\psi_r(\tau, \xs)$ and the contribution originating from
the initial condition $\psi_0(\tau, \xs)$, where
$\psi_r(0,\xs)$ is initially zero and $\psi_0(0,\xs)=1$. In the communal
recombination model, the term describing the production of novel recombinants reduces to $\rs\Phi(\tau)$ and the two
contributions evolve according to:
\begin{eqnarray}
\nonumber
\partial_\tau \psi_0(\tau, \xs)&=&\gr \psi_0-\epsilon \psi_0^2\\ 
\label{eq:psi_r}
\partial_\tau \psi_r(\tau, \xs)&=&r\Phi(\tau)+\gr\psi_r-\epsilon
(2\psi_0\psi_r +\psi_r^2) \ . 
\end{eqnarray}
Recombination enters the equation for $\psi_0$ only through a reduction of the growth rate and the solution for $\psi_0(\tau, \xs)$ is simply
\begin{equation}
\label{eq:asex}
\psi_0(\tau, \xs) = \frac{e^{\gr^2/2}}{\epsilon \int_0^{\tau}d\tau' e^{\gr'^2/2} +
e^{(\gr-\tau)^2/2}} 
\end{equation}
The integral $\Phi_0(\tau)=\int d\xs P(\xs,\tau)\psi_0(\tau, \xs)$ acts as a source for $\psi_r(\tau, \xs)$. The initial behavior of $\Phi_0(\tau)$ is quite different for $\epsilon\gg 1$ and $\epsilon \ll 1$ and we will analyze these cases separately below.

\subsection{Survival probability of neutral mutations}
\label{sec:app_neutral}
Here, we present the steps necessary to arrive at the result for the survival probability (Eq.~\ref{eq:Pnx}).
The survival probability is given by the generating function of $\Pnumber(n,\tau)$, evaluated at $\lambda=0$. In our rescaled variables, the survival probability therefore corresponds to $\Phi(\tau)$ with $\epsilon=\sigma^{-1}\gg 1$. The long time solution for $\Phi(\tau)$ given in Eq.~\ref{eq:longtime} contains the free parameters $\ti$ and $\Thi$, defined by matching with the short time solution which depends explicitly on the initial condition (see above). To establish the matching, we trace the short time solution for $\epsilon\gg 1$ through several regimes until the initial condition is forgotten. At small times $\tau\ll 1$, we have
\begin{equation}
\Phi(\tau)\approx\Phi_0(\tau) = \frac{1}{\epsilon \tau +1 } 
\end{equation}
This rapid initial decay corresponds to the early loss of the allele due to neutral processes. Selection starts to matter only after $\tau\sim 1$ yielding an accelerated decay of  $\Phi_0(\tau)$ 
\begin{equation}
\Phi_0(\tau) =   \frac{1}{\epsilon\sqrt{2\pi}} e^{-\frac{\tau^2}{8}-\frac{1}{2}\rs\tau-\rs^2/2}
\end{equation}
The Gaussian decay of the survival probability of un-recombined alleles is due to the increasing mean fitness, i.e.~the adapting population wave moving past the genotype the mutation initially arose on. Hence $\Phi_0(\tau)$ does not contribute to the long time behavior, which is dominated by the part of the solution driven by recombination. We therefore have to calculate how much weight is transferred from $\psi_0$ to $\psi_r$ via the term $\rs\Phi(\tau)$ in Eq.~\ref{eq:psi_r}. In the case $\epsilon\gg 1$, the dominant balance for $\psi_r(\tau, \xs)$ is $\rs\Phi_0(\tau)\approx 2\epsilon \psi_r(\tau, \xs)\psi_0(\tau, \xs)$, resulting in an initially steady $\Phi_r(\tau) = \frac{\rs}{2\epsilon}$. After the initial condition has decayed and $\Phi_0(\tau)\ll \Phi_r(\tau)$ the solution $\psi_r(\tau, \xs)$ acquires the form of Eq.~\ref{eq:linear_solution} with the only memory of the initial condition in $\Phi_r(\tau)$. The crossover time is obtained by equating $\Phi_0(\tau)$ with $\Phi_r(\tau)$, which yields $\ti=\sqrt{8\log (2/\rs)}-2\rs$. Using this $\ti$ and $\Thi \approx \sqrt{-4\log \rs }$ (comp.~Eq.~\ref{eq:linear_solution}), we can match the short time solution to the long time solution in Eq.~\ref{eq:longtime}.

\begin{figure}[htp]
\begin{center}
  \includegraphics[width=0.32\columnwidth]{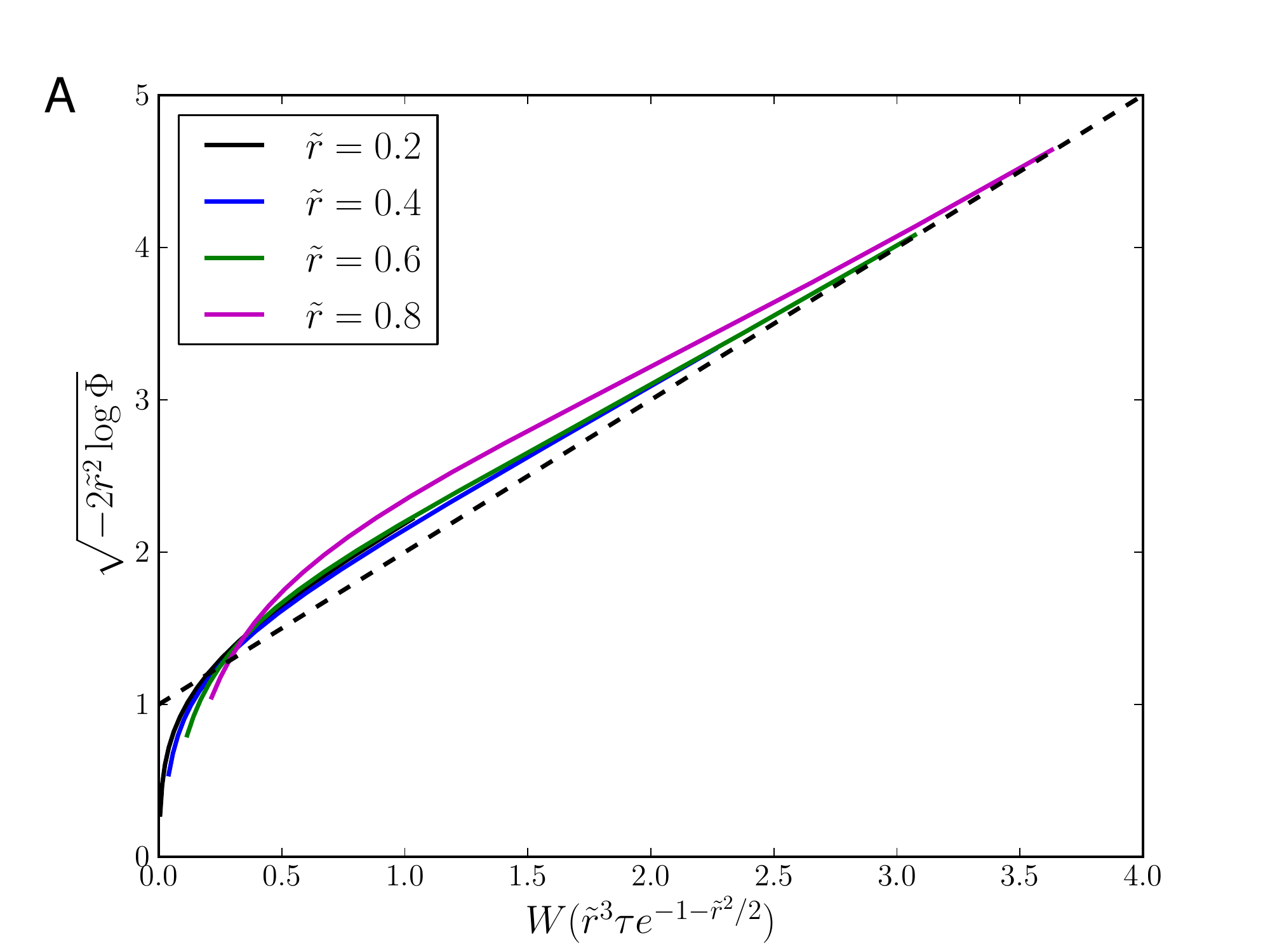}
  \includegraphics[width=0.32\columnwidth]{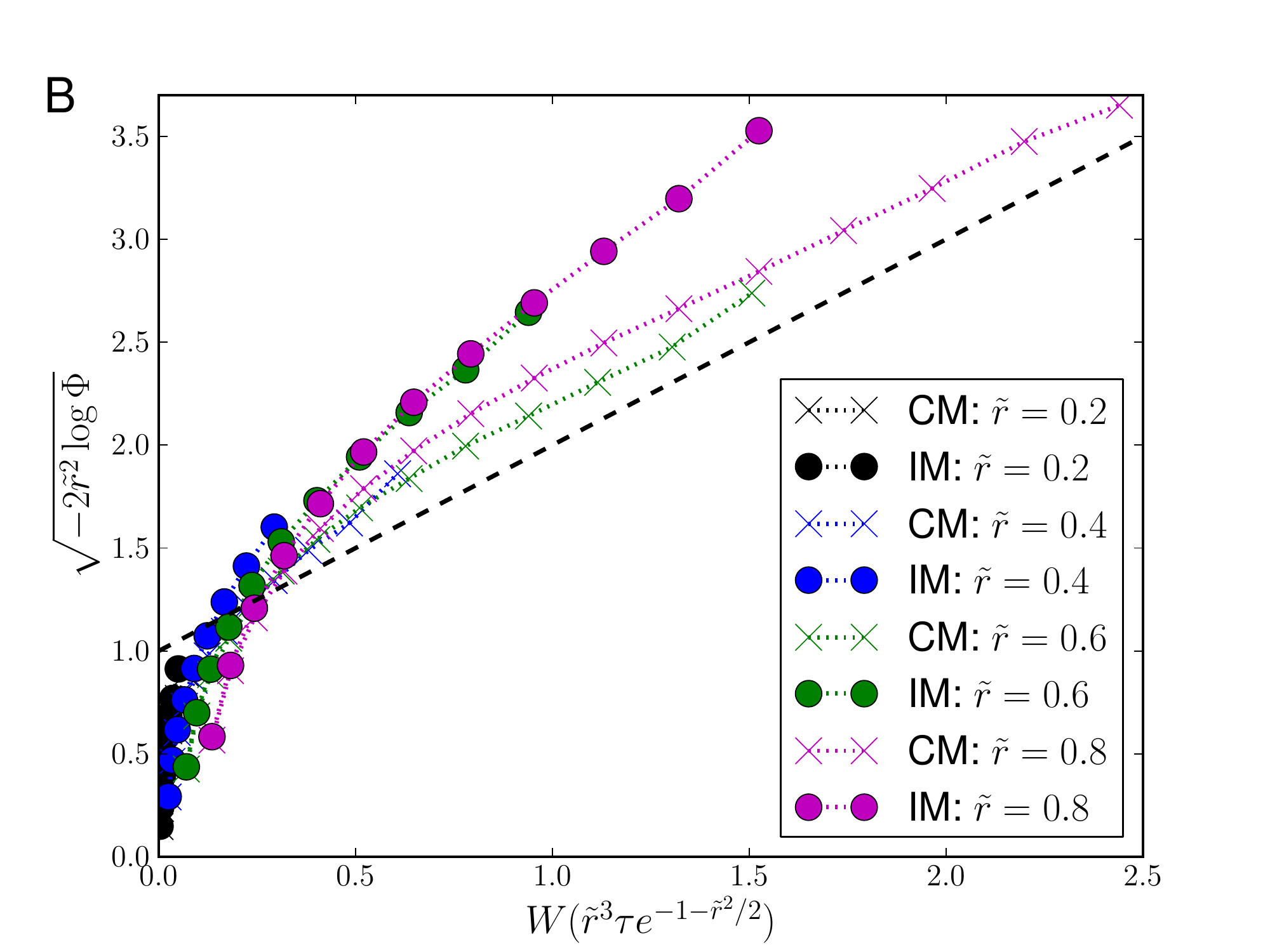}
  \includegraphics[width=0.32\columnwidth]{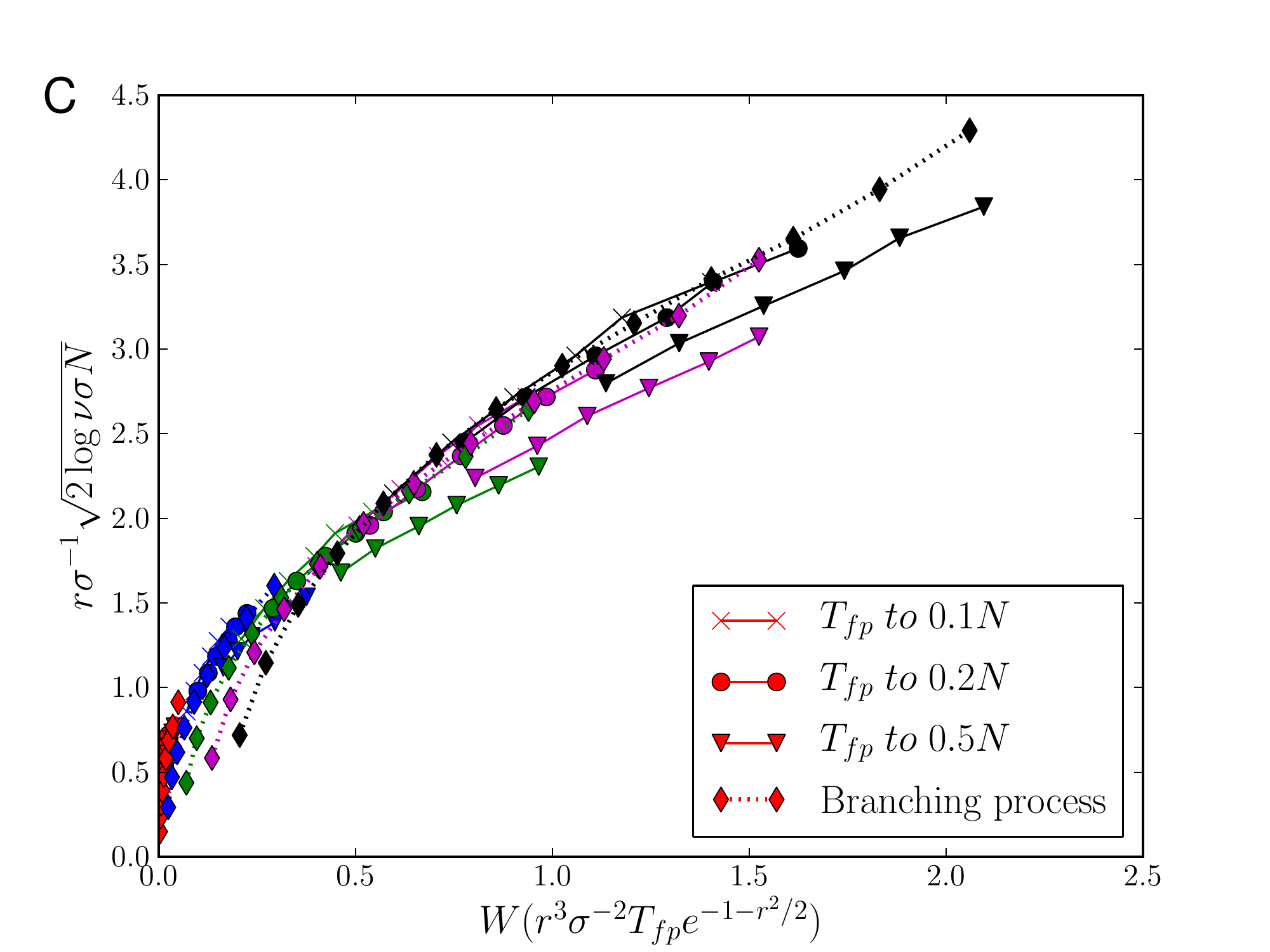}
  \caption[labelInTOC]{Asymptotic behavior of the survival probability for the communal and infinitesimal recombination model. Panel A: The survival probability $\Phi(\tau)$, obtained by numerical solution of Eq.~\ref{eq:phi} for the communal recombination model, exhibits the predicted asymptotic behavior $\sqrt{-2\rs^2 \log \Phi}\sim W(\rs^3 \tau e^{-1-\rs^2/2})$ (Eq.~(\ref{eq:Pnx})). Panel B: The survival probability in the infinitesimal model (IM) and the communal model (CM), determined by simulating the branching process, exhibit a similar scaling collapse, indicating that the dependence on parameters is similar. Panel C: The first passage time, $\Tfp$, to allele copy number $\nu N$,  is well described by the branching process approximation for the dynamics of alleles in a finite population as long as the frequency $\nu \ll 1$. This is confirmed here by plotting $\sqrt{2\rs^2\log \sigma \nu N}$ vs.~$W(\rs^3 \sigma\Tfp e^{-1-\rs^2/2})$. The collapse of data for different $\rs$ and different $\nu$ confirms the predicted parameter dependence, Eq.~(\ref{eq:Tfix}), while the concordance with the branching process results confirms the branching process model. Deviation from the branching process prediction are apparent for $\nu=0.5$. Recombination rates in panel C follow the same color code as in panels A\&B, while black symbols correspond to $\rs=1$.}
  \label{fig:Pnx_asymptotic}
  \label{fig:Pnx_model_compare}
\end{center}
\end{figure}

To illustrate the validity of the approximations made in the analysis we solved Eq.~\ref{eq:psi} numerically. 
The left panel of Fig.~\ref{fig:Pnx_asymptotic} shows the square root of the logarithm of the rescaled survival probability, $\sqrt{-2\rs^2\log\Phi(\tau)}$, plotted as a function of $W(\rs^3 \tau e^{-1-\rs^2/2})$. Data for different outcrossing rates collapse onto the same curve, indicating that the expression in Eq.~(\ref{eq:longtime}) captures the dependence on $\rs$.  

The above analysis assumed the communal recombination model, i.e.~a model where the offspring fitness independent of the parental fitness. To demonstrate that the communal recombination model captures the behavior of the infinitesimal model, we compare results obtained from stochastic simulations of the branching process for the two models in Figure \ref{fig:Pnx_model_compare}. The infinitesimal model exhibits the same scaling collapse, indicating that the dependence on the parameters is similar. The two models do however show quantitative differences, which can be interpreted as less efficient recombination in the infinitesimal model. This is expected, as recombination in the infinitesimal model only halves the correlation with either parent, while this correlation is completely destroyed in the communal recombination model. Note, however, that the dependence of $\Pnx$ on $\rs$ is strong, so that a rescaling of $\rs$ substantially changes the absolute numbers.

\subsection{Beneficial and deleterious mutations}
\label{sec:app_bene_dele}
To analyze the behavior of $\Phi(\tau)$ in the case of $\dss\neq 0$, we track back to Eq.~\ref{eq:Phidot}, focusing on the case where $\rs\Thc\gg1$.
\begin{equation}
\partial_\tau \Phi(\tau) = (\dss -\rs e^{-\rs\Thc}) \Phi(\tau)
\end{equation}
This equation highlights the fact that as long as $\dss \ll \rs e^{-\rs\Thc}$, the dynamics of $\Phi(\tau)$ is dominated by fluctuations and is therefore similar to the neutral case. For a negative $\dss$, the decay of $\Phi(\tau)$ is accelerated, while a positive $\dss$ slows down and eventually halts the decay of $\Phi(\tau)$, see Figure \ref{fig:quasineutrality}. The crossover to the behavior dominated by the intrinsic fitness effect occurs when the two terms equal. Comparing with Eq.~\ref{eq:non_extinction_rate} yields the crossover condition 
\begin{equation}
\dss \approx \frac{\log(\rs^3 \tau )}{\rs^2 \tau } \quad \mathrm{or}\quad \tau\approx \frac{\log(\rs/\dss )}{\rs^2 \dss }
\end{equation}
Before this crossover, the effect of selection on $\dss$ is a simple attenuation or amplification of $\Phi(\tau) = e^{\dss \tau} \Psi(\tau)$, where $\Psi(\tau)$ is approximately the neutral solution, with the slight difference that the crossover point $\Thc \approx \sqrt{-2\log (\epsilon \rs \Psi(\tau))-2\dss\tau}$. The correction to the crossover point is still small when selection on the intrinsic fitness effect $\ds$ starts to dominate the dynamics of $\Phi(\tau)$.

\subsection{Distribution of the allele copy number}
\label{sec:app_copy_number}
The probability $p(n,T)$ of finding $n$ copies of the allele after time $T$ can be calculated from the generating function $\hat{p}(\lambda, T)$ by contour integration in the complex $\lambda$-plane:
\begin{equation}
\label{eq:contour_integral}
\Pnumber(n,\tau) =\frac{1}{2\pi i}\oint_\mathcal{C}d\lambda \frac{\hat{p}(\lambda, T)}{\lambda^{n+1}} 
\end{equation}
The contour $\mathcal{C}$ has to encircle the origin and no other singularities of $\hat{p}(\lambda, T)$. For $n>0$, we can evaluate the contour integral by deforming the contour to run alongside the singularities away from zero, which are tightly controlled by $\lambda^{-n}$ for $n\gg 1$. To do so, we have to study the analytic properties of $\hat{p}(\lambda, T)$ or equivalently of $\Phi(\tau) = \sigma^{-1}\epsilon^{-1}(1-\hat{p}(\lambda, T))$, where $\epsilon=1-\lambda$.
For negative $\epsilon=1-\lambda$ and $s\geq 0$, $\psi(\xs, \tau)$ has a singularity at $\Thc$, the value separating the regime of small $\gr$ where the linear equation is valid and the saturated regime where $\psi(\tau)=\gr/\epsilon$, comp.~Eq.~\ref{eq:linear_solution}. $\Phi(t)$ is the integral of $\psi(\tau, \xs)$  over $\xs$ and therefore has a branch cut for negative real $\epsilon$. Otherwise, $\Phi(\tau)$ is an analytic function of $\epsilon$. We can use this branch cut to evaluate the contour integral in Eq.~\ref{eq:contour_integral}.
Substituting $\Phi(\tau)$ into Eq.~\ref{eq:contour_integral} in changing the integration variable to $\epsilon$, we have
\begin{equation}
\Pnumber(n,\tau) =\frac{\sigma}{2\pi i}\oint_\mathcal{B}d\epsilon \frac{\epsilon \Phi(\tau)}{(1-\epsilon)^{n+1}} \approx \frac{\sigma}{2\pi i}\oint_\mathcal{B}d\epsilon e^{\log \epsilon +n\epsilon+\log\Phi(\tau)}
\end{equation}
where $\mathcal{B}$ is the contour encircling the branch cut. For large $n$ the dominant contributions to the integral come from $\epsilon \ll 1$ and we have to find a solution for $\Phi(\tau)$ for small $\epsilon$. 

To this end, we back-track to Eq.~\ref{eq:asex} and integrate the evolution equation for $\Phi(\tau)$ for $\epsilon\ll 1$. We find 
\begin{equation}
\Phi_0(\tau)=\begin{cases}
              e^{-\rs\tau} &\tau<\sqrt{-2\log\epsilon}\\
               \frac{1}{\sqrt{\pi\epsilon\tau}}e^{-\frac{\tau^2}{8}-\frac{1}{2}\rs\tau-\rs^2/2} & \tau>\sqrt{-2\log\epsilon}
             \end{cases}
\end{equation}

For intermediate $\rs$, the matching between the short term solution and the long time solution occurs at $\ti=\Thi=\sqrt{-2\log \epsilon}$, after which we have 
\begin{equation}
\Phi(\tau) = \frac{\Thc}{\epsilon \rs} e^{-\Thc(\tau)^2/2} \quad with \quad \Thc(\tau)\approx
\rs^{-1}\log (\rs^3(\tau-\ti)+e^{r\Thi}) 
\end{equation}
For large $\Thi$, $\Thc(\tau)\approx \Thi + \rs^2(\tau-\ti)e^{-\rs\Thi}$ and the branchcut integral becomes
\begin{equation}
\Pnumber(n,\tau) \approx \frac{\sigma}{2\pi \rs i}\oint_\mathcal{B}d\epsilon \Thi e^{ne^{-\Thi^2/2} -\Thi^2/2 - \Thi\rs^2(\tau-\ti)e^{-\rs\Thi}}
\end{equation}
The exponent of the integrant peaks when 
\begin{equation}
\label{eq:branchcut_saddle}
n\Thi e^{-\Thi^2/2} + \Thi - \rs^2(\rs\Thi-1)(\tau-\ti)e^{-\rs\Thi}=0
\end{equation}
For large $n$, the dominant balance is between the first term of Eq.~(\ref{eq:branchcut_saddle}) and the remainder, which requires that $n\approx e^{\Thi^2/2}$ and hence $\epsilon \approx n^{-1}$. Using $\Thi=\ti=\sqrt{2\log n}$, the asymptotic copy number distribution becomes
\begin{equation}
\label{eq:Pn_asymptotics}
\Pnumber(n,\tau)\sim\frac{1}{n\sqrt{2 \log n}}e^{-\frac{\left(\Thi+ \rs^{-1}\log(1+\rs^3(\tau-\ti)e^{-\rs\Thi})\right)^2}{2}} \ .
\end{equation}
The leading dependence of $\Pnumber(n,\tau)$ on $n$ is $\Pnumber(n,\tau)\sim n^{-2}$. Note, however, that the crossterm from the exponent yields another subdominant factor $e^{\alpha \sqrt{\log n}}$. By contrast, the corresponding expression in the high recombination regime is $\Pnumber(n,T)\sim T^{-2}e^{-n/T}$.

The allele frequency spectrum, $\Pallele(n)$, i.e.~the probability to find $n$ copies of a derived allele irrespective of its age, is given by the time average of $\Pnumber(n,\tau)$ over mutations that arose at different times $\tau$ in the past. It depends on how quickly the survival probability decays, as well as on the shape of $\Pnumber(n,\tau)$. For large $n$ one can expand the logarithm in the exponent of Eq.~(\ref{eq:Pn_asymptotics}) and integrate $\Pnumber(n,\tau)$ over $\tau$  to obtain
\begin{equation}
\Pallele(n)=\int_{0}^{\infty} d\tau\; \Pnumber(n,\tau)
 \sim \begin{cases}
   \frac{e^{\rs\sqrt{2\log n}}}{n^2\log n} & \rs \ll 1\\
   n^{-1} & \rs\gg 1
   \end{cases}
\end{equation}
Even though $\Pnumber(n,\tau)$ decays much faster for $\rs \gg 1$ (exponential) than for $\rs \ll 1$ (algebraically), the allele frequency spectrum is steeper for $\rs \ll 1$ than for $\rs\gg 1$. The long tail of the latter is due to the contribution of very old alleles with very flat exponential $\Pnumber(n,\tau)$. For $\rs \ll 1$, the clonal amplification and rapid extinction of most alleles give rise to steep allele frequency spectra.

\subsection{The double mutation probability: Stochastic tunneling}
\label{sec:app_tunneling}
To calculate tunneling probabilities \citep{Iwasa:2004p32851,Weissman:2009p23398}, we need to know the distribution of $w=\int_0^t dt' n(t')$ treated as a stochastic variable, i.e.~the size of transient ``bubbles'' of mutant individuals in a large population.  In analogy to Eq.~\ref{eq:BME}, the distribution $p(w,T|k,t,\x)$ for a mutation being present on background $\x$ in $k$ copies obeys the following equation
\begin{equation}
\begin{split}
-(\partial_t -k\partial_w) p(w,T|k,t,\x)=& -k(B+D+\rec)p(w,T|k,t,\x)
+ k B p(w,T|k+1, t,\x)+k D p(w,T|k-1,
t,\x)\\&+\rec k\int dw' \int dx' K_{\x\x'} p(w-w',T|k-1,t,\x)p(w',T|1,t,\x')
\end{split}
\end{equation}
Instead of the generating function, we now consider the Laplace transform 
$\hat{p}(z,T|k,t,\x)=\int dz e^{-zw}p(w,T|k,t,\x)$. Thanks to the convolution property of Laplace transforms and the fact that the fate of the $k$ alleles present at time $t$ is independent of each other we have: $\hat{p}(z,T|k,t,\x)=\hat{p}^k(z,T|1,t,\x)$. As before, we scale variables by $\sigma^{-1}$ and times by $\sigma$. The equation for $\phi(\tau,z,\x) = \sigma^{-1}(1 - \hat{p}(z,T|1,t,\x))$ then is 
\begin{equation}
\label{eq:comoving_cumulative_n}
\partial_\tau \phi(\tau,z,\xs) = z
+\rs\Phi(\tau, z)+\gr\phi(\tau,z,\xs)-\phi^2(\tau,z,\xs) \ ,
\end{equation}
where $\gr = \xs + \tau +\dss +z-\rs$. At $\tau=\sigma(T-t)=0$, $w=0$ and therefore $\phi(\tau,z,\xs)=0$. If tunneling is rare and happens only on time scales longer than the life time of ``bubbles'', we do not need to know the full time-dependent solution for $\phi(\tau,z,\xs)$ but can send $\tau$ to infinity. 
The above equation can again be solved by matching the asymptotic solutions at large negative and positive $\gr$ which in the steady state read
\begin{equation}
\phi(z,\xs) \approx \begin{cases}
				e^{\frac{\gr^2}{2}} [z+\rs\Phi(z)] & \gr\ll\Thc\\
\gr & \gr\gg\Thc
\end{cases}
\end{equation}
The crossover point $\Thc$ is determined by $e^{\frac{\Thc^2}{2}} [z+\rs\Phi(z)] =\Thc$. 
The ``solvability condition" obtained by integrating Eq.~\ref{eq:comoving_cumulative_n} w.r.t.~Gaussian distribution of $\xs$ yields:
\begin{equation}
\label{eq:solvability_z}
0 = z
+\dss\Phi(z) -\int {d\xs \over \sqrt{2 \pi}} e^{-\xs^2/2} \phi^2(z, \xs)\ ,
\end{equation}
For small $\rs$, the integral on the right hand side is dominated by the vicinity of the crossover point $\Thc$. Combining with the matching condition, one finds for 
%
%
$\Phi(z)$
\begin{equation}
\label{eq:steady_state}
\Phi(z) = z \ { 1-e^{-\rs \Thc} \over \rs  e^{-\rs\Thc}-s}
\end{equation}
Substituting into the crossover condition we obtain an equation for $\Thc$:
\begin{equation}
\label{eq:self_consist}
 z \ {\rs -\dss \over \rs e^{-\rs \Thc } -\dss } =
\Thc e^{-\frac{\Thc^2}{2}}
\end{equation}
In the neutral case ($|\dss|\ll \rs e^{-\rs \Thc}$), the location of the crossover is at $\Thc=\sqrt{-2\log (ze^{\rs \Thc }/\Thc)}$ which implies for $\Phi(z)$
\begin{equation}
\Phi(z) = z \ { e^{\rs \Thc} -1 \over \rs  } = \begin{cases}
                                                     z \sqrt{-2\log z} & \rs\Thc\ll 1 \\
                                                     ze^{\rs \Thc}/\rs & \rs \Thc\gg 1
                                                      \end{cases}
\end{equation}
In case the intrinsic effect of the mutation dominates the denominator of Eq.~\ref{eq:self_consist} ($|\dss| \gg \rs e^{-\rs\Thc}$) one finds
\begin{equation}
\label{eq:deleterious_limit}
\Phi(z) = \frac{z}{|\dss|} ( 1-e^{-\rs \Thc} )=\begin{cases}
                                                        \frac{z\rs\sqrt{-2\log z}}{|\dss|} & \rs\Thc\ll 1\\
                                                        \frac{z}{|\dss|} & \rs\Thc\gg 1
                                                        \end{cases}
\end{equation}
The result for $\Phi(z)$ given by Eq.~\ref{eq:steady_state} and its different limits is directly relevant to the calculation of the probability of secondary events, e.g.~an additional mutation as discussed in the earlier in \emph{Stochastic tunneling and complex adaptations}.  

For completeness we also calculate the probability distribution of $w=\int_0^{\infty} dt n(t)$ which is given by the inverse Laplace transform of $\Phi(z)$:
\begin{equation}
P(w) =\oint_\mathcal{C} \frac{dz}{2\pi i} e^{zw}\bar{\Phi}_z
\end{equation}
where $\mathcal{C}$ is the contour encircling the branch cut. For example in the intermediate asymptotic  $\rs \Thc \ll 1$ regime when $\Phi(z) =  z  \sqrt{  \log z^{-2} }$, we have
\begin{eqnarray}
&P(w) = \oint_\mathcal{C} \frac{dz}{2\pi i} e^{zw} z  \sqrt{  \log z^{-2} }=\int_0^{\infty} \frac{dy}{\pi } e^{-yw} y  \ \mathrm{Im} [ \sqrt{  2\log y^{-1}+i2\pi } ]\\
&\approx \int_0^{\infty} {dy y e^{-yw} \over    \sqrt{  \log y^{-2}}} \sim {w^{-2} \over    \sqrt{  \log w^{2}}}
\end{eqnarray}
where we used an approximation valid for $w\gg 1$.
This can be compared to the ``draft-less" result, obtained also in the $\rs \gg 1$ limit: in that case $\Phi(z) = \sqrt{z}$, inverse Laplace transform of which gives $P(w) \sim w^{-3/2}$
\citep{Weissman:2009p23398}.

\section{Computer simulation methods}
\label{sec:app_model_implementation}
\emph{Fisher-Wright simulation.} To efficiently simulate the dynamics of large populations, we keep track of classes of individuals with identical genomes, which are encoded by bitstrings. The fitness $\x$ of all individuals of a class is simply the sum of the contributions from all loci. In our discrete generation scheme, a pool of gametes is produced to which each class contributes a Poisson distributed number of gametes with mean $c\exp(-(\x-\bar{\x}+\alpha))$. Here, $c$ is the size of the class, $\bar{\x}$ is the mean fitness in the population and $\alpha = 1-N/\bar{N}$ keeps the population size $N$ approximately constant at $\bar{N}$. Fitness defined as growth rate in the continuous time model naturally assumes this exponential form in a discrete generation model, simply by integration over one generation. 

To implement facultative mating, a fraction $\rec$ of the gametes are paired up and from each pair two offspring are produced by assigning at random the genes of the parents to either offspring. The remaining $1-\rec$ fraction of gametes is copied into the next generation without recombination, i.e.~they have gone through an asexual reproduction cycle. 

The genome of each class has a fixed length $L=1024$ and new mutations are introduced at a locus whenever this locus is monomorphic, i.e.~whenever a previous mutation either went extinct or fixed. For each monomorphic locus, an individual is chosen at random and the mutation is introduced in this single individual. This scheme of keeping a fixed number of loci polymorphic has the advantage of making optimal use of the computational resources, i.e.~keep every locus polymorphic while maintaining an overall low mutation rate. The mutation rate, however, becomes a dependent quantity which fluctuates around a value that depends on other parameters of the population: $N\mut_b$ is simply the average number of loci that become monomorphic in one generation, see \citep{Neher:2010p30641}. For large $L$, the number of mutations introduced in each generation is much greater than one and doesn't fluctuate greatly. 

To establish a steadily adapting population, $80\%$ of the loci are kept polymorphic with beneficial mutations with effect size $\ds_0$ which each generation is rescaled such that the overall fitness variance is $\sigma^2=0.0025$. This rescaling is done to be able to specify $\rec/\sigma$ and $\ds/\sigma$ explicitly. Since fluctuations of $\sigma$ in large populations with many polymorphic loci are small, this does not change the properties of the dynamics and the same results are obtained letting $\sigma$ be freely determined by the population. The remaining 20\% of the loci are used to study the fate of mutations with fixed effect size $\ds$. Measurements are performed after an equilibration period of $20000$ generations at intervals of one hundred generations. The source code is available from the authors on request.

\emph{Simulation of the branching process.} The simulation of the population dynamics is complemented by a  simulation of the branching process which can be directly compared to analytic calculations. We simulate the process described by Eq.~(\ref{eq:BME}) using an event driven algorithm \citep{Gillespie:1977p35367}. The simulation keeps track of all individuals that currently carry the allele in question. For each individual, the time $t+\Delta t$ of the next event is determined by drawing a $\Delta t$ from an exponential distribution with parameter $B(t)+D+\rec$, where $B$ and $D$ are the birth and death rates, respectively. At time $t+\Delta t$, a birth, death, or recombination event are performed with probabilities proportional to $B(t+\Delta t)$, $D$, and $\rec$, respectively. In case of death, the individual is deleted, in case of birth it is duplicated with the exact same fitness, in case of recombination, its fitness is redrawn according to the recombination kernels in Eqs.~(\ref{eq:fullmodel}) or (\ref{eq:poolmodel}). (Note that the waiting time distribution for the next event is not exactly exponential, since the birth rate is time dependent. This, however, amounts only to a correction of order $\sigma^2\ll 1$.) 
For the recombination function of the communal model, we need not simulate Eq.~(\ref{eq:BME}) but can solve numerically for the generating function of the $\Pnumber(n,T,t)$, given in Eq.~(\ref{eq:phi}). To this end, we discretize the fitness $\x$ and solve for the vector $\psi(\tau,\x)$ using the ODE solver of SciPy \citep{Oliphant:2007p25672}. The results obtained through stochastic simulation of the communal model agree with the numerical solution for the generating function, as it has to be. 

\bibliography{bib}

\newpage
\section{Supplement}
\author{R.~A.~Neher${}^{*}$}
\author{B.~I.~Shraiman${}^{*\ddagger}$}
\maketitle
\subsection*{Derivation of the backward Master equation for $\Pnumber(n,T)$}
The fundamental quantity of the branching process is the probability distribution, $p(n,T)$, of observing $n$ copies of the allele $T$ generations after it originated. A backward equation for $\Pnumber(n,T)$ can be derived by considering the probability $\Pnumber(n,T | k,t,\x)$ of having $n$ copies at time $T$, given their were $k$ copies on a background with fitness $\x$ at time $t$. 
$\Pnumber(n,T | k,t-\Delta t,\x)$ now can be expressed as a sum over possible intermediate states at time $t$:
\begin{equation}
\begin{split}
\Pnumber(n,T | k,t-\Delta t,\x)=& \Pnumber(n,T | k,t,\x)(1-\Delta tk(2+x-\bar{x}(t)+s+r)) 
\\& + \Delta t\left[k(1+x-\bar{x}(t)+s)\Pnumber(n,T | k+1,t,\x)+k\Pnumber(n,T | k-1,t,\x)\right] \\
& + \Delta t rk\sum_{n'}\int_{x'}K_{xx'} \Pnumber(n-n',T | k-1,t,\x)\Pnumber(n',T | 1,t,\x') \ ,
\end{split}
\end{equation}
where $\bar{x}(t)$ is the mean fitness of the population at time $t$. 
The different terms have straightforward interpretations: The first term is the probability that nothing happens in the small time interval $\Delta t$, the second accounts for a division of one of $k$ individuals, which happens with rate $k(1+x-\bar{x}(t)+s)$, the third term accounts for the death of one of the $k$, while the last term accounts for outcrossing of one of the $k$, producing a new individual with fitness $\x'$ and removing one with fitness $\x$. The outcrossing term is then summed over all possible ways the $k-1$ individuals with fitness $\x$ and the one with fitness $\x'$ can give rise to $n$ individuals at time $T$. Sending $\Delta t$ to zero and rearranging terms results in an ODE for $\Pnumber(n,T | k,t,\x)$
\begin{equation}
\begin{split}
-\partial_t \Pnumber(n,T | k,t,\x)=& -\Pnumber(n,T | k,t,\x) k(2+x-\bar{x}(t)+s+r)
\\&+k(1+x-\bar{x}(t)+s)\Pnumber(n,T | k+1,t,\x)+k\Pnumber(n,T | k-1,t,\x)\\
&+rk\sum_{n'}\int_{x'}K_{xx'} \Pnumber(n-n',T | k-1,t,\x)\Pnumber(n',T | 1,t,\x')
\end{split}
\end{equation}
This is equation 1 from the main text.

\subsection*{Derivation of the equation for the generating function }
To remove the convolution over $n$ it is convenient to consider the generating function $\hat{p}(\lambda,T|k,t,\x)=\sum_n \lambda^n \Pnumber(n,T|k,t,\x)$. Multiplying the above equation by $\lambda^n$ and summing over $n$ yields 
\begin{equation}
\begin{split}
-\partial_t \hat{p}(\lambda,T|k,t,\x)=&
-k(2+x-\bar{x}(t)+s+r)\hat{p}(\lambda,T|k,t,\x) 
\\&+k(1+x-\bar{x}(t)+s)\hat{p}(\lambda,T|k+1, t,\x)+k\hat{p}(\lambda,T|k-1, t,\x)
\\&+rk\int_{x'}K_{xx'}\hat{p}(\lambda,T|k-1,t,\x)\hat{p}(\lambda,T|1,t,\x)
\end{split}
\end{equation}
All $k$ initial individuals are independent, hence
$\hat{p}(\lambda,T|k,t,\x)=\hat{p}^k(\lambda,T|t,\x)$ and the right hand side 
is $-\partial_t \hat{p}^k(\lambda,T|t, \x)=-k
\hat{p}^{k-1}(\lambda,T|t,\x) \partial_t \hat{p}(\lambda,T|t,\x)$.  We can therefore divide the equation by $\hat{p}_x^{k-1}(\lambda,T|t,\x)$  to obtain
\begin{equation}
\begin{split}
-\partial_t \hat{p}(\lambda,T,t, \x)=&
-(2+x-\bar{x}(t)+s+r)\hat{p}(\lambda,T,t, \x)  +
(1+x-\bar{x}(t)+s)\hat{p}^2(\lambda,T,t, \x)+1
+r\int_{x'}K_{xx'}\hat{p}(\lambda,T,t, \x)
\end{split}
\end{equation}
The generating function has the boundary condition
$\hat{p}(\lambda,T|T,\x)=\lambda$, which follows from $\Pnumber(n,T|k,T,\x)=\delta_{nk}$.
Substituting  $\hat{p}(\lambda,T| t,\x)=1-\phi(\lambda,T, t,\x)$ removes the constant term.
\begin{equation}
\begin{split}
\partial_t \phi(\lambda,T, t,\x)=&-r\int_{x'}K_{xx'}\phi(\lambda,T, t,\x')-(x-\bar{x}(t)+s-r)\phi(\lambda,T, t,\x)+(1+x-\bar{x}(t)+s)\phi^2(\lambda,T, t,\x)
\end{split}
\end{equation}
which now has boundary condition $\phi(\lambda,T, T,\x)=1-\lambda$.
Assuming that selection is weak on the timescale of one generation ($\sigma\ll 1$), we
can approximate $1+x-\bar{x}(t)+s$ by 1 and arrive at Eq.~18 of the main text:
\begin{equation}
\begin{split}
\label{eq:phi}
-\partial_t \phi(\lambda,T, t,\x)=
r\int_{x'}K_{xx'}\phi(\lambda,T, t,\x')+(x-\bar{x}(t)+s-r)\phi(\lambda,T, t,\x)-\phi^2(\lambda,T, t,\x)
\end{split}
\end{equation}

\subsection*{Solution for $\Phi(\tau)$ at low $\rs$}
In the main text, we presented a solution to the two equations (Eqs.~(21) and (23))
\begin{equation}
\label{eq:supp_linear_solution}
\psi(\tau, \xs)=\begin{cases} 
             re^{\gr^2/2}\int_0^\tau
d\tau' \Phi(\tau')e^{-\gr'^2/2} & \gr<\Thc \\
	\gr/\epsilon & \gr>\Thc
             \end{cases}
\end{equation}
and 
\begin{equation}
\label{eq:supp_solvability}
\partial_\tau
\Phi(\tau) = \dss\Phi(\tau) - \epsilon\int d\xs P(\tau, \xs)\psi^2(\tau, \xs)
\end{equation}
in a regime of intermediate $\rs$ ($1/\sqrt{\log N}<\rs<1$). An additional solution with qualitatively different properties exists at low $\rs$. While the assumption of a Gaussian fitness distribution is questionable in this range of $\rs$, we nevertheless present the solution here for completeness. Proceeding as before, we evaluate the integral in Eq.~(\ref{eq:supp_linear_solution}) by expanding exponent around its maximum.  The maximum is located at $\tau' = \tau-\gr - \alpha(\tau')$ where $\alpha(\tau)= -\Phi(\tau)^{-1}\partial_\tau \Phi(\tau)$.
Assuming $\alpha(\tau)$ changes slowly with time, we have
\begin{equation}
\psi(\tau)\approx\sqrt{2\pi}
\rs\Phi(\tau)e^{\frac{(\gr+\alpha)^2}{2}}
\end{equation}
This solution is valid below $\Thc$, where $\psi(\tau)$ crosses over the
linear saturated form $\gr/\epsilon$. In addition to the matching condition
at $\Thc$, we use Eq.~\ref{eq:supp_solvability} for $\ds=0$ to determine $\alpha$:
\begin{equation}
\label{eq:supp_Phidot}
\partial_\tau\Phi(\tau) = -\alpha\Phi(\tau) =  - \rs\Phi(\tau)e^{\Thc(\alpha-\rs)+\alpha^2/2-\rs^2/2}
\end{equation}
For $\rs\Thc\gg 1$ we recover the solution with small $\alpha$ given in the main text. For smaller $\rs$, however, we find $\alpha(\tau)\approx \frac{\log \rs}{\Thc}$. Solving the matching condition for $\Thc$ and differentiating with respect to $\tau$ yields $\partial_\tau \Thc= -\frac{1}{\Thc}\alpha(\tau)$. This is readily solved for this case of $\Thc\rs< 1$
\begin{equation}
\label{eq:supp_longtime_theta}
\Thc(\tau)= 
(-3(\tau-\ti) \log \rs + \Thi^3)^{1/3} \ .
\end{equation}
Substituting this solution for $\Thc(\tau)$ into the expression for the rescaled generating function yields
\begin{equation}
\label{eq:longtime}
\Phi(\tau)= \frac{\Thc}{\epsilon\rs}e^{-\frac{(\Thc+\alpha)^2}{2}} \sim
e^{-\frac{\left[-3(\tau-\ti) \log
\rs + \Thi^3\right]^{2/3}}{2}}
\end{equation}
Hence in this low $\rs$ regime, the decay of the survival probability is qualitatively different from the regime of intermediate $\rs$.

\subsection*{Effective clone-based model}
To rationalize the behavior of the continuous time branching process, we considered the following simplified model discussed in the main text: Genotypes expand clonally and produce recombinant offspring. The offspring start growing simultaneously in the next ``effective" generation after all clones from the previous generation have disappeared. The relevant quantity now is the number of clones or distinct genotypes, rather than the number of individuals. To understand the dynamics of the number of clones, we need to know how many clones a single clone can produce. 

Consider a single genotype with background fitness $\chi$ which is carrying a mutation of effect size $\dss$. The expected number of recombinant offspring from this genotype is $\xi=\rs\int_0^\infty dt n_\chi(t)$, where $n_\chi(t)$ is the copy number trajectory. The Laplace transform $\hat{p}(\xi)$ of $p(\xi)$ obeys the equation ($\phi(z) = 1- \hat{p}(z)$)
\begin{equation}
\partial_\chi \phi(\chi, z) = \rs z +  (\chi+\dss-\rs(1-z))\phi(\chi, z) - \phi(\chi,z)^2 \ ,
\end{equation}
which is a simpler version of the Eq.~(42) (main text) since only one single clone is considered and recombination to daugther clone is ignored. This equation can be solved asymptotically in the regimes of large and small $\chi-\rec$.
\begin{equation}
\phi(\chi, z) = \begin{cases}
             \rs z e^{(\chi+\dss-\rs(1-z))^2/2}\int_{-\infty}^{\chi+\dss-\rs(1-z)} dx e^{-x^2/2}
             & \chi-\rs\ll \Theta_c \\
             (\chi-\rs(1-z)) & \chi-\rs\gg \Theta_c
             \end{cases}
\end{equation}
where $\Theta_c\approx \sqrt{-2\log \rs z}$. The initial fitness of the genotype, $\chi$,
is Gaussian distributed and the Laplace transform has to be averaged over $\chi$.
\begin{equation}
\label{eq:avg_laplace}
\phi(z)=\int_{-\infty}^{\infty} \frac{d\chi}{\sqrt{2\pi}}e^{-\chi^2/2}\phi(\chi, z)
\end{equation}
Let's look at the mean number $\langle \xi \rangle$ of recombinant offspring first, which is given by differentiating with respect to $z$ and setting $z=0$, which in turn sends $\Theta_c$ to infinity. Integration by parts yields
\begin{equation}
\begin{split}
\langle \xi \rangle =\partial_z \int_{-\infty}^{\infty}
\frac{d\chi}{\sqrt{2\pi}}e^{-\chi^2/2}\phi(\chi) &=
\rs \int_{-\infty}^{\infty}
\frac{d\chi}{\sqrt{2\pi}}e^{-\chi (\rs-\dss) +r^2/2}\int_{-\infty}^{\chi-\rs+\dss}
dx e^{-x^2/2} 
 = \frac{\rs}{\rs-\dss}
\end{split}
\end{equation}
Hence the mean number of recombinant offspring is 1 for a neutral mutation and approximately $1+\dss/\rs$ for mutations with small effect. To evaluate the integral in Eq.~(\ref{eq:avg_laplace}) at finite $z$, we have to account for the cross-over of $\phi(\chi, z)$ at $\chi-\rs = \Theta_c \approx \sqrt{-2\log \rs z}$. For small $z$, the cross-over translates into a cut-off of the integral. Again, the integral can be evaluated by parts: 
\begin{equation}
\begin{split}
\phi(z) 
&=\int_{-\infty}^{\Theta_c+\rs-\dss}
\frac{d\chi}{\sqrt{2\pi}} \rs z e^{-\chi (\rs(1-z)-\dss) +
\rs^2(1-z)^2/2}\int_{-\infty}^{\chi-\rs(1-z)+\dss} dx e^{-x^2/2} +
\int_{\Theta_c+\rs-\dss}^\infty \frac{d\chi}{\sqrt{2\pi}} e^{-\chi^2/2}(\chi -\rs(1-z))
\\
&\approx \frac{\rs z}{\rs-\dss}\left[1-e^{-\Theta_c (\rs-\dss) -
\rs^2/2}\right]
\end{split}
\end{equation}
Hence, the generating function of the average number of recombinant offspring generated by a random genotype is given by 
\begin{equation}
\label{eq:bubblelaplace}
\hat{p}(z) = 1-\phi(z) \approx 1-\frac{\rs z}{\rs-\dss}\left[1-e^{-\Theta_c (\rs-\dss)}\right] \ ,
\end{equation}
where the last expression is valid for $z\ll 1$ and $\rs\ll 1$.

The actual number $m$ of recombinant offspring (novel clones) generated by one clone is Poisson distributed with mean $\xi$. The generating function of $m$ is therefore
\begin{equation}
\sum_m \lambda^m P(m) = \sum_m \lambda^m \int_0^\infty d \xi\ p(\xi) e^{-\xi}
\frac{\xi^m}{m!} = \int_0^\infty d \xi\ p(\xi) e^{-\xi(1-\lambda)} =
\hat{p}(1-\lambda) = 1-\phi(1-\lambda)
\end{equation}
We will now use this result to calculate how the number of clones that descend from a particular genotype evolves over time.

\subsubsection*{The stochastic dynamics of the number genotypes}
As long as the clones we are tracking constitute a small fraction of the population, different clones are independent. The probability to go from $k$ to $m$ clones in one effective generation has therefore the generating function $\hat{P}(\lambda, k)=\hat{P}^k(\lambda, 1)=(1-\phi(1-\lambda))^k$.
To study the dynamics of the number of clones over many generations, we need to know how this propagator behaves when iterated.
\begin{equation}
\begin{split}
&\sum_m \lambda^m \sum_{m_1, \ldots m_n} P(m, m_n)P(m_n, m_{n-1})\ldots P(m_1,
k) = \sum_{m_n} \hat{P}^{m_n}(1-\lambda)\sum_{m_1, \ldots m_{n-1}} P(m_n,
m_{n-1})\ldots P(m_1, k)\\
&= \left[\hat{P}(1-\hat{P}(1-\hat{P}(1-\ldots \hat{P}(1-\lambda))))\right]^k = \left[1-\phi\circ
\phi\ldots \circ \phi(1-\lambda)\right]^k = \left[1-\Phi_n(1-\lambda)\right]^k
\end{split}
\end{equation}
Using the result for the Laplace transform in Eq.~(\ref{eq:bubblelaplace}), we arrive at the difference equation
\begin{equation}
\Phi_{n+1} - \Phi_n \approx \left[\dss/\rs-e^{-\rs{\sqrt{-2\log \rs\Phi_n}}}\right]\Phi_n 
\end{equation}
This is exactly the differential equation we have derived using the
continuous time mode when the effective generation time is set to $\rs^{-1}$. The latter is reasonable since the $\rs^{-1}$ is the turnover time by recombination.

\subsection*{Supplementary Figure: Population distribution}
\begin{figure}[htp]
\begin{center}
  \includegraphics[width=0.7\columnwidth]{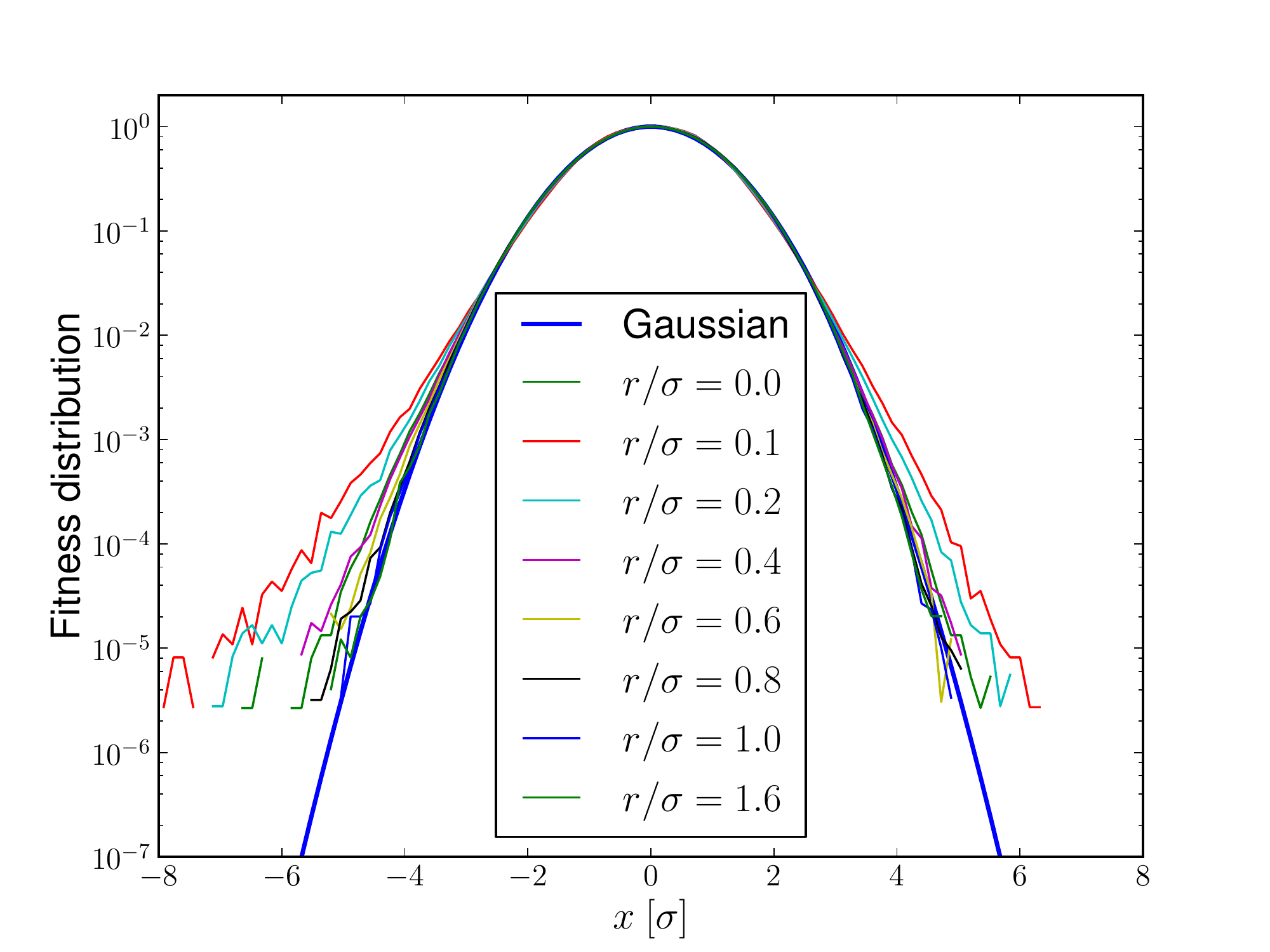}
  \caption[labelInTOC]{The fitness distribution of a population adapting at many loci is Gaussian. The figure shows the fitness distribution measured using our computational model for different ratios $\rec/\sigma$. Only at $\rec/\sigma=0.1$ or 0.2, one sees (slight) deviations from the Gaussian. Our analysis applies in the range $1>\rec/\sigma>1/\sqrt{\log N}$, where the fitness distribution is Gaussian all the way into the tails.}
  \label{fig:supp_gauss}
\end{center}
\end{figure}

\subsection*{Supplementary Figures: Background Selection}

\begin{figure}[htp]
\begin{center}
  \includegraphics[width=0.45\columnwidth]{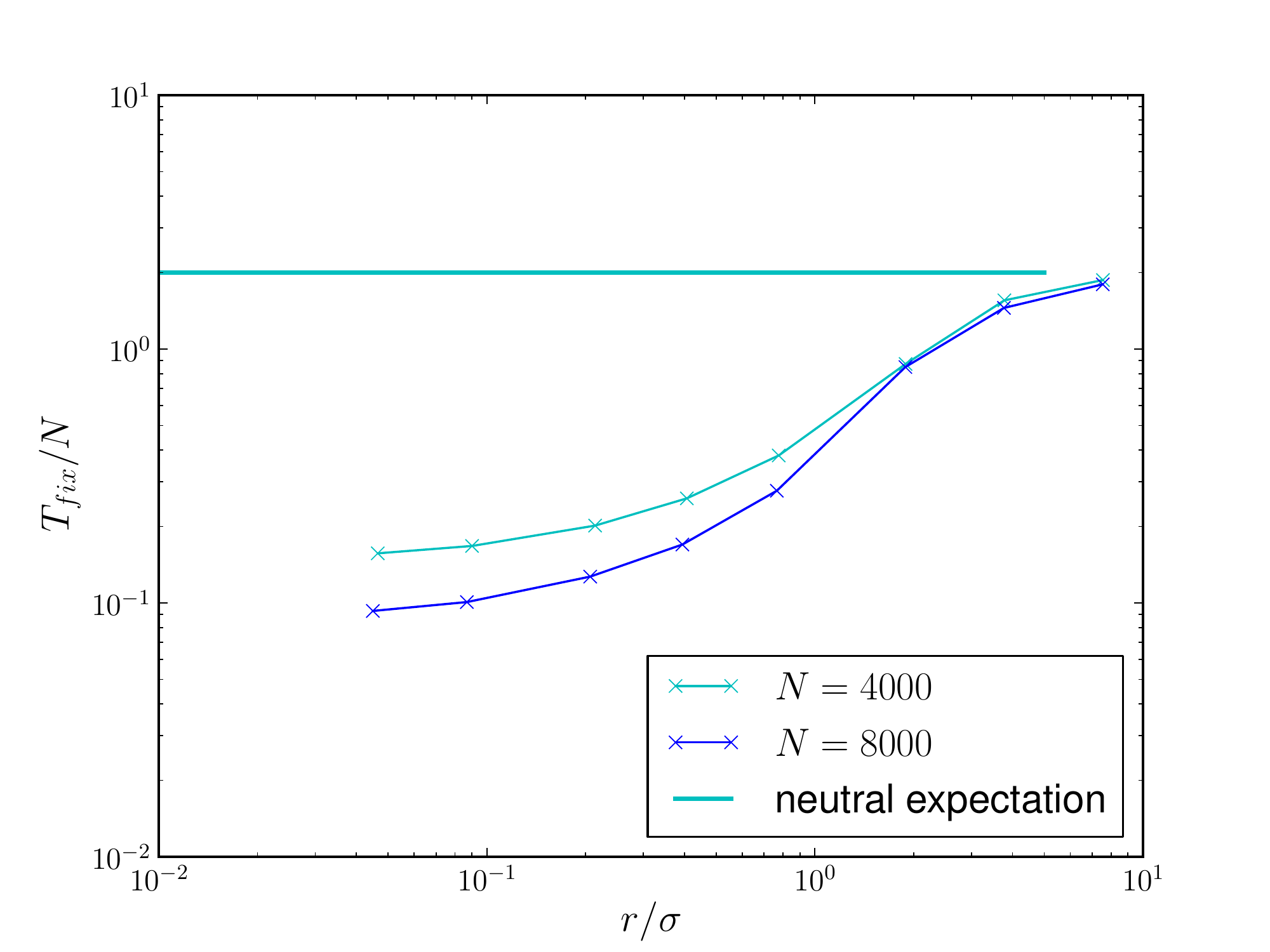}
  \includegraphics[width=0.45\columnwidth]{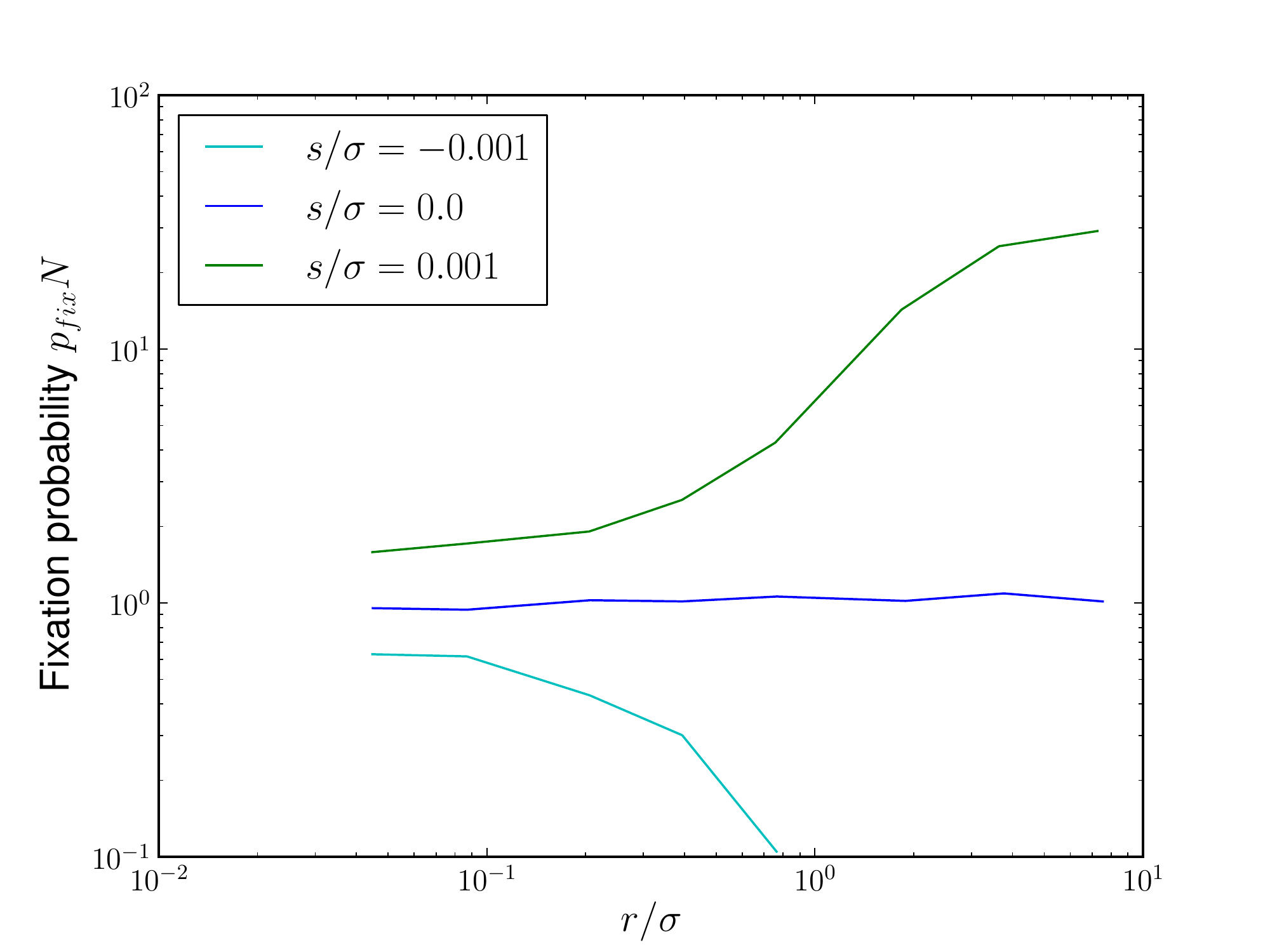}
  \caption[labelInTOC]{Fixation times and fixation probabilities with background selection, in analogy to Fig.~2 of the manuscript. The left panel shows the mean fixation time of neutral mutations in populations of different sizes at different ratios of $\rec/\sigma$. Fixation times are normalized by $N$. In contrast to Fig.~2 of the manuscript, the fitness variation here does not result from sweeping beneficial mutations but from many deleterious mutations. Comparing Fig.~2 to this figure, one sees that the effect of background selection on the fixation time of neutral mutations is very similar to that of multiple sweeps in a facultatively sexual population. The right panel shows the fixation probability of mutations with different effects on fitness for different ratios $\rec/\sigma$, normalized to the neutral expectation $N^{-1}$. Again, the fitness variance is due to background selection rather than sweeps, but the effect on the fixation probability is similar. These observations are consistent with the argument that from the perspective of a novel mutation, the nature of the fitness variation is irrelevant. What matters is the dynamics of the fitness of individual genotypes relative to the mean: For background selection, the fitness of genotypes declines due to accumulation of deleterious mutations, while in the case of adaptation the mean increases steadily. }
  \label{fig:supp_purifying}
\end{center}
\end{figure}

\begin{figure}[htp]
\begin{center}
  \includegraphics[width=0.45\columnwidth]{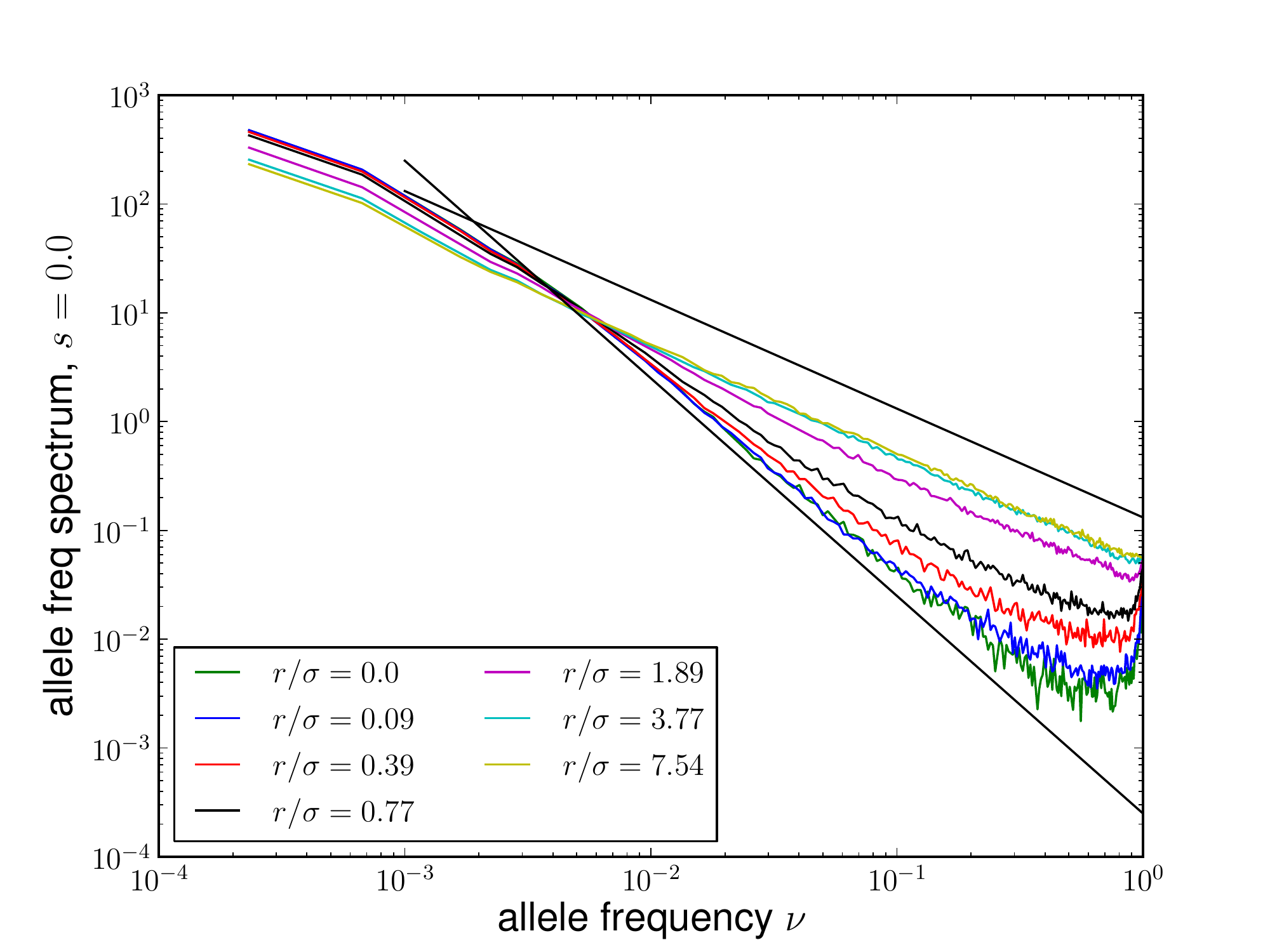}
  \includegraphics[width=0.45\columnwidth]{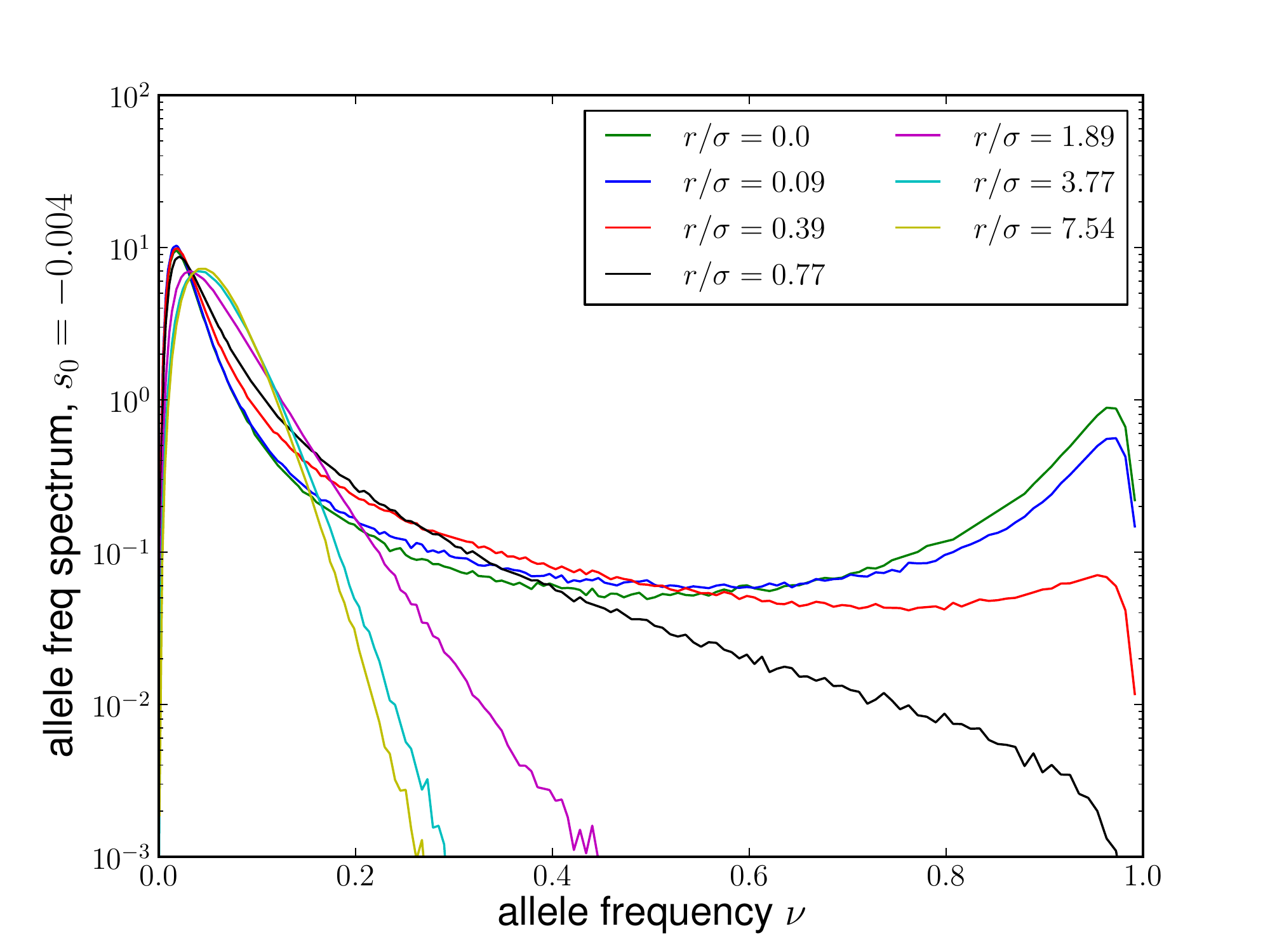}
  \caption[labelInTOC]{Allele frequency spectra of neutral (left) and deleterious mutations (right) in a background selection scenario, in analogy to Fig.~3b of the manuscript. At low recombination rates ($\rec/\sigma<1$), the frequency spectrum of neutral mutations falls off much more rapidly than expected in a neutral model, very similar to what is observed for the scenario of continuous adaptation. The predicted behavior $\sim\nu^{-2}$ is indicated by the steeper black line. Only when $\rec/\sigma\gg 1$ does the spectrum agree with the neutral prediction ($\sim\nu^{-1}$ indicated by upper straight black line). The right panel shows the frequency spectrum of the deleterious mutations responsible for the fitness variation with effect size $s_0=-0.004$. At low recombination rates, allele frequencies are close to fixation, either in the bad ($\nu=1$) or the good ($\nu=0$) state. At high recombination rates, the allele frequency spectra are distributed around their equilibrium value $\nu=\mu/s_0=0.0625$.}
  \label{fig:supp_purifying}
\end{center}
\end{figure}

\end{document}